\documentclass[aps,prd,natbib,onecolumn,groupedaddress,nofootinbib,preprintnumbers,superscriptaddress,nolongbibliography,secnumarabic]{revtex4-1}  
\usepackage{graphicx}
\usepackage{amsmath}
\usepackage{amsfonts}
\usepackage{bm,bbm}
\usepackage[colorlinks=true,citecolor=red,linkcolor=blue,breaklinks=true]{hyperref}
\usepackage[figure, table]{hypcap}
\usepackage{amssymb}
\usepackage{xcolor}
\usepackage{subfigure}
\usepackage{setspace}
\usepackage{footnote}
\usepackage{multirow}
\usepackage[normalem]{ulem}
\usepackage[utf8]{inputenc}
\usepackage{mathrsfs}
\usepackage{slashed}
\usepackage{longtable}
\usepackage{enumitem}
\usepackage{cancel}
\usepackage{booktabs}
\usepackage{blindtext}
\usepackage{listings}
\usepackage{float}
\usepackage{topcapt}
\usepackage{cleveref}
\usepackage{makecell}
\usepackage[table,usenames,dvipsnames]{xcolor}

\usepackage{hhline}

\usepackage{orcidlink}
\newcommand{\orcid}[1]{\,\orcidlink{#1}}

\usepackage{xcolor}
\definecolor{linkcolor}{rgb}{0.0, 0.47, 0.75}
\definecolor{citecolor}{rgb}{1.0, 0.5, 0.0}
\hypersetup{
  linkcolor  = linkcolor,
  citecolor  = linkcolor,
  urlcolor   = linkcolor,
  colorlinks = true
}

\definecolor{blue-violet}{rgb}{0.54, 0.17, 0.89}

\definecolor{olive}{rgb}{0.6, 0.8, 0.4}

\definecolor{byzantine}{rgb}{0.74, 0.2, 0.64}

\definecolor{random}{rgb}{0.3, 0.3, 0.74}

\makeatletter

\newcommand{\sectionnotoc}[1]{%
    \refstepcounter{section}%
    \begingroup
    \let\addcontentsline\@gobblethree
    \section*{\thesection\quad #1}%
    \endgroup
}

\newcommand{\subsectionnotoc}[1]{%
    \refstepcounter{subsection}%
    \begingroup
    \let\addcontentsline\@gobblethree
    \subsection*{\thesubsection\quad #1}%
    \endgroup
}

\makeatother

\begin{document}
\preprint{CERN-TH-2026-250}
\preprint{IFT-UAM/CSIC-26-144}

\title{\vspace*{1cm}Collider and direct detection as key probes of asymmetric dark matter:\\ Signals from a compressed Inert Higgs doublet}

\author{\vspace*{5mm}David Cerde\~no \orcid{0000-0002-7649-1956}}
\email{davidg.cerdeno@ift.csic.es}
\affiliation{Instituto de F\'isica Te\'orica IFT-UAM/CSIC, Cantoblanco, E-28049, Madrid, Spain}

\author{Miguel Escudero Abenza\orcid{0000-0002-4487-8742}}
\email{miguel.escudero@cern.ch}
\affiliation{Theoretical Physics Department, CERN, 1 Esplanade des Particules, CH-1211 Geneva 23, Switzerland}

\author{Pietro Ghedini \orcid{0009-0001-4861-4867}}
\email{pietro.ghedini@ific.uv.es}
\affiliation{Instituto de Física Corpuscular (IFIC), CSIC‐UV and Departament de Física Teòrica, UV, Spain}

\author{Guilherme Guedes \orcid{0000-0003-2881-9483}}
\email{guilherme.guedes@cern.ch}
\affiliation{Theoretical Physics Department, CERN, 1 Esplanade des Particules, CH-1211 Geneva 23, Switzerland}

\author{Thomas Hambye \orcid{0000-0003-4381-3119}}
\email{thomas.hambye@ulb.be}
\affiliation{Service de Physique Th\'eorique, Université Libre de Bruxelles, Bld du Triomphe, CP 225, 1000 Brussels, Belgium}

\author{Abel Muñoz-Navarro \orcid{0009-0009-0436-8148}}
\email{amunoz@icc.ub.edu}
\affiliation{Departament de Física Quàntica i Astrofísica (FQA) and Institut de Ciències del Cosmos (ICCUB), Universitat de Barcelona (UB),  c.~Martí i Franquès, 1, 08028 Barcelona, Spain}

\author{Elena Pressendo \orcid{0009-0001-9629-6090}}
\email{elena.pressendo2@unibo.it}
\affiliation{Dipartimento di Fisica e Astronomia, Università di Bologna, via Irnerio 46, 40126 Bologna, Italy}

\begin{abstract}
\vspace*{3mm}\noindent 

There is a cosmic imbalance of particles over antiparticles in the visible sector of the Universe and the same may hold true for dark matter. Many models for asymmetric dark matter have been proposed to date and a particularly appealing set is the one where the dark asymmetry is directly inherited from the baryon asymmetry. In this study we thoroughly investigate the signatures of the minimal realization of this idea. In this model a new (stable) inert Higgs, $H'$, plays the role of the mediator that facilitates the asymmetry transfer from the visible to the dark sector. It also constitutes a small dark matter sub-component,
while the dominant component stems from a singlet scalar asymmetry. A combination of phenomenological and cosmological constraints requires this new inert Higgs doublet to be quite light $46\,{\rm GeV}<m_{H'}< 580\,{\rm GeV}$ and its spectrum rather compressed, $\Delta M = m_{H^\pm}-m_{H_0} \lesssim 1\,{\rm GeV}$ and $\delta m = m_{A_0}-m_{H_0}\lesssim 5\,{\rm MeV}$. Therefore its charged component can be relatively long-lived on collider length-scales $10^{-2}\,{\rm cm}\lesssim c \tau \lesssim 10^2\,{\rm cm}$. By recasting searches for compressed Higgsinos at the LHC, we derive new limits on this particle and show that, while at present most of the parameter space for this state is still unconstrained, large and key regions of parameter space will be tested at the HL-LHC by a combination of searches for disappearing charge tracks and displaced soft tracks accompanied with large missing energy. Quite remarkably, the lifetime of the charged component of the doublet can be correlated with the dark matter interaction rate at direct detection experiments and we show that the next generation of direct detection experiments has a unique sensitivity to the parameter space in the model. The fact that a small neutral mass splitting is a prediction of the asymmetric setup makes it an interesting candidate to account for the high energy nuclear recoil event recently reported by the LZ experiment. More generally, our detailed phenomenological study holds for any dark matter model involving an inert Higgs doublet with mass below 580 GeV.
\end{abstract}

\maketitle
\begingroup
\setlength{\parskip}{0.5pt}
\hypersetup{linkcolor=black}
\makeatletter
\def\l@f@section{%
  \addpenalty{\@secpenalty}%
  \addvspace{0.4em}%
}
\makeatother
\tableofcontents
\endgroup

\section{Introduction}\label{sec:intro}

There is no evidence for antimatter structures in the Universe~\cite{Steigman:1976ev,Cohen:1997ac}, and the fact that today there are about $10^9$ photons per baryon   implies that there must have been a primordial imbalance between baryons and antibaryons in the early Universe at the level of one part in a billion~\cite{Kolb:1990vq,Rubakov:2017xzr}. As the primordial Universe cooled down to $T\simeq 40\,{\rm MeV}$, all antibaryons quickly annihilated away and only baryons remained in the Universe. There are many good ideas for how this primordial imbalance between baryons and antibaryons may have been generated~\cite{Kolb:1990vq,Rubakov:2017xzr}, and prime to them is the leptogenesis scenario, which elegantly emerges from the very well-motivated seesaw mechanism that naturally explains the observed neutrino masses and mixings~\cite{Fukugita:1986hr}. 

Given that we observe an imbalance of particles over antiparticles in the visible sector, one wonders whether a similar situation takes place in the dark sector, in particular since the dark matter and baryon energy densities are so similar, $\Omega_{\rm DM}/\Omega_{b}\sim 5$. This is not a new idea and it has been thoroughly explored in the literature, see e.g.~\cite{Hut:1979xw,Nussinov:1985xr,Gelmini:1986zz,Barr:1990ca,Kaplan:1991ah} for some of the early references, 
and~\cite{Cirelli:2024ssz,Zurek:2013wia,Petraki:2013wwa,Boucenna:2013wba,Davoudiasl:2012uw} for reviews. 
In this context, a particularly appealing class of models are those where a dark matter asymmetry is not generated on its own but where it is instead transferred from the asymmetry that must be present in the visible sector, see e.g.~\cite{Kaplan:2009ag,Cohen:2009fz,Cui:2011qe,Ibe:2011hq,Servant:2013uwa,Boucenna:2015haa,Dhen:2015wra,Escudero:2025dei,Blennow:2026psy}. These scenarios are very interesting for several reasons: 1) they provide a connection between the dark matter abundance and the baryon asymmetry of the Universe; 2) they do not require new sources of CP or baryon number violation beyond those at the origin of the Standard Model baryogenesis (as the visible-to-dark particle asymmetry transfer can be CP-conserving); 3) the dark matter particles involved cannot be arbitrarily heavy, $m\lesssim \mathcal{O}(10)\,{\rm TeV}$~\cite{Griest:1989wd}, as for the dark matter asymmetry to dictate the dark matter abundance these particles need to annihilate faster than usual WIMPs. This makes these scenarios testable, because the asymmetry transfer will take place, typically, at a temperature of order the dark matter particle mass.

In this context, Ref.~\cite{Escudero:2025dei} recently developed what may be the simplest phenomenologically viable model within the general class of scenarios discussed above. It is simple because it is renormalizable and requires only the presence of two new states beyond the Standard Model: $\phi$, a complex singlet scalar, and $H'$, a new inert Higgs doublet. In this scenario, an asymmetric $\phi$ population represents the dominant dark matter component and these particles need to be somewhat heavy, $m_\phi \gtrsim 700\,{\rm GeV}$. The inert Higgs doublet, $H'$, plays a threefold role in the model: 1) since it carries both Standard Model and dark sector quantum numbers, it is the portal that facilitates the transfer of the visible asymmetry into the dark matter; 2) being lighter than $\phi$, it constitutes the dominant final state to which $\phi \bar{\phi}$ particles annihilate efficiently in the early Universe. This is critical as dark matter particles which efficiently annihilate into Standard Model Higgs doublets are now ruled out by direct detection experiments~\cite{EscuderoAbenza:2025cfj}; 3) the lightest component of the $H'$ doublet is stable and therefore also constitutes a dark matter component but, being lighter than $\sim 580\,{\rm GeV}$~\cite{Bottaro:2022one,Cirelli:2005uq}, it is a subdominant one (as in this case its annihilation into SM particles is very efficient and the $H'$ relic density is below the observed dark matter one). Quite generically, the lighter the $H'$, the more subdominant it is and the more asymmetric the dark matter is.

The phenomenology of this asymmetric dark matter model turns out to be outstandingly rich and its study is the purpose of this work. Since the lightest neutral component of the $H'$ doublet is part of the dark matter, direct detection experiments can test its interaction rate with the Standard Model Higgs. In the absence of strong cancellations, the very strong limits from direct detection experiments on elastic spin-independent interactions~\cite{LZ:2024zvo,XENON:2025vwd,PandaX:2024qfu} lead to the conclusion that the neutral and charged components must be very degenerate in mass. In addition, the $Z$ boson mediates potentially very efficient inelastic scatterings between the neutral $H'$ components and this represents also another key signature of the model. This is particularly interesting in the context of the recently observed high energy nuclear recoil event reported by LZ~\cite{LZ:2026axp} because the rate is the same as for Higgsinos. Chiefly, unlike for those models~\cite{Fan:2026kxx} a small mass splitting, $\delta m \sim 400\,{\rm keV}$, is perfectly natural and furthermore is required to be smaller than $\delta m \sim 5\,{\rm MeV}$ for the mechanism of Ref.~\cite{Escudero:2025dei} to operate successfully. Together with the possible scatterings from the $\phi$ particle, the model can lead to three distinct signatures at direct detection experiments. Given the motivation for this scenario, the potential dark matter interpretation of the signal seen by LZ and the order of magnitude improvements in the sensitivity of direct detection experiments in the years to come, it is timely to explore the signals that this model can lead to.

\begin{figure}[t]
    \centering
    \includegraphics[width=1.00\linewidth]{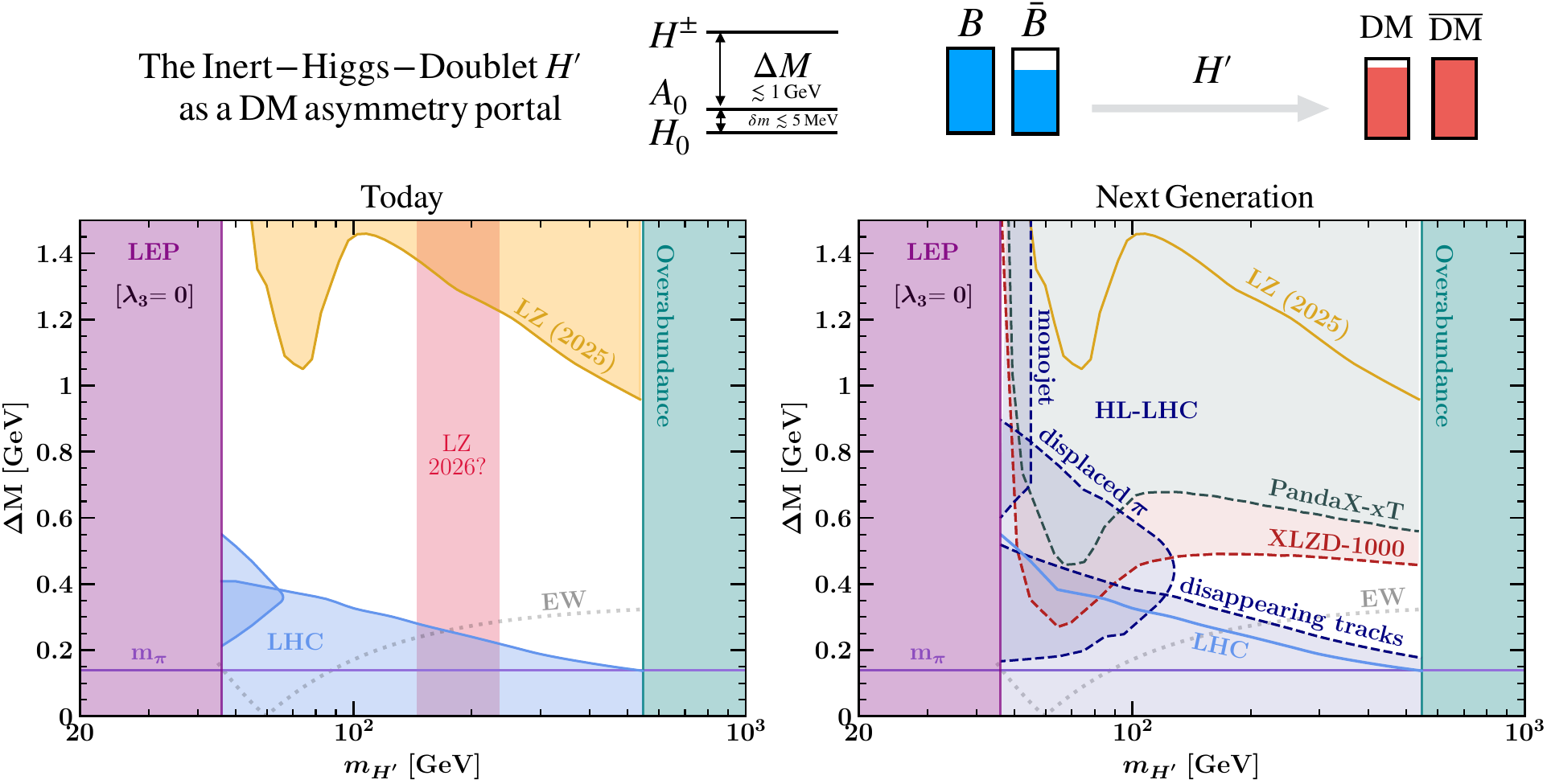}
    \caption{Parameter space for a compressed inert Higgs doublet $H'$ with mass $m_{H'}$. \textit{Left:} Current constraints from direct detection and the LHC. The upper region is excluded by LZ (90\% CL) while the LHC contours (95\% CL) show the limits from both disappearing charge tracks and from displaced pion searches which turns out to only be effective at $m_{H'}\lesssim 60\,{\rm GeV}$. \textit{Right:} Sensitivities from the next generation of direct detection experiments and the HL-LHC. We have chosen $\lambda_3 = 0$ (see Section~\ref{sec:DD}) and the EW dotted line represents the mass splitting as generated by electroweak interactions. In the left panel we highlight where in the presence of a boosted velocity distribution on Earth the inert Higgs may explain the LZ signal while being in agreement with the null observations from high energy neutrinos from the sun, see Section~\ref{sec:DD_inelastic}. Irrespectively of this, it is clear that a light inert Higgs doublet is rather unconstrained at present but has true discovery potential in the next years via a combination of searches at direct detection experiments and at the HL-LHC. Only the small white region in the bottom right corner of the right panel would remain untested.}
    \label{fig:bigDelta_EWcorr_Money}
\end{figure}

Critically, a $H'$ with a rather compressed mass spectrum can also lead to striking signatures at the LHC such as disappearing charged tracks or displaced soft tracks. Analyses of these signatures have been performed by the ATLAS and CMS collaborations in searches for degenerate Higgsinos~\cite{CMS:2023mny,ATLAS:2026hnb,ATLAS:2025lhc,CMS:2026ias,ATLAS:2024umc}. In this work, we recast them to the inert doublet case, identifying the regions of parameter space that can be targeted by collider experiments both with current data and considering the future high-luminosity (HL) phase of the LHC. It turns out that these signals are very complementary to searches at dark matter direct detection experiments, as is evident in Fig.~\ref{fig:bigDelta_EWcorr_Money}. This figure highlights the present constraints (left) and future sensitivities on $H'$ (right), and summarizes some of the most important results of our study. As direct detection probes larger $\Delta M$ (the mass splitting between the charged and neutral components of $H'$), disappearing track searches target the region from below, as smaller splittings give rise to longer tracks in the detector. The richness of signatures and yet its currently weak bounds directly follow from the compressed nature between the components of the $H'$ doublet, such that it can still be as light as 46 GeV despite having an electrically charged $\pm 1$ component, leading to a real discovery potential both at the LHC and in direct detection experiments.

More generally, beyond the key role of $H'$ as a portal to asymmetric dark matter, an inert Higgs doublet is interesting on its own right as a dark matter candidate~\cite{Barbieri:2006dq,LopezHonorez:2006gr,Hambye:2009pw,Arina:2009um}. Our phenomenological study holds for any scenario involving a light stable inert Higgs doublet $H'$, so that it applies to a wide class of models and scenarios, see~\cite{Blinov:2015qva,Belyaev:2016lok,Belanger:2021lwd} for previous studies along these lines.

This paper is structured as follows: in Section~\ref{sec:model} we review the main components of the model and highlight its key features including various cosmological considerations. In Section~\ref{sec:DD} we consider the direct detection signatures of the scenario, and in particular those of the inert Higgs doublet. We show that it can explain the high energy event by LZ only if one assumes momentum distributions of the dark matter particles crossing the Earth that are boosted for high momentum values, which allows to evade the limits on neutrinos from the sun. In Section~\ref{sec:collider} we investigate the collider phenomenology and in particular recast searches for long-lived electroweakly interacting states at the LHC. We conclude in Section~\ref{sec:global} by putting the results from all the previous sections together and show that most of the parameter space will soon be tested by a combination of HL-LHC and direct detection searches.

\section{Minimal Genesis of Asymmetric Dark Matter via an Inert Higgs Doublet}\label{sec:model}

As stressed in the introduction, the phenomenological study we perform in this work applies to any model containing an inert Higgs doublet lighter than 580 GeV. In this regime, its relic density is totally fixed by electroweak interactions (thus only constituting a sub-component of dark matter), and in the absence of strong cancellations its components need to be rather close in mass (hence compressed) to comply with limits from dark matter direct detection. Before discussing this phenomenology, in this section we would like to motivate such a situation on the basis of the recently proposed simple asymmetric dark matter setup of Ref.~\cite{Escudero:2025dei}, which precisely requires the presence of such a light inert dark matter component with a compressed spectrum.

Dark matter may be composed of different numbers of particles and antiparticles, and scenarios where this imbalance is directly linked to the baryon asymmetry of the Universe are particularly interesting. Arguably, the simplest models are those in which an asymmetry in the Standard Model baryonic sector is generated (for instance, through leptogenesis) and this asymmetry is later transferred to the dark matter sector, as those set ups do not require any new sources of CP or B/L violation beyond those at the origin of baryogenesis. Several models along these lines have been considered but most of them either suffer from severe constraints from direct detection experiments or the LHC or suffer some potential tunings, such as the requirement of extremely tiny (asymmetry erasing) ${\rm DM^2}$ terms, allowed in dark matter singlet setups based on a $Z_2$ symmetry. In this context, the model of Ref.~\cite{Escudero:2025dei} is a new, perfectly viable, simple, UV complete, and theoretically appealing scenario. It works as follows.

If a baryon asymmetry was generated at temperatures above $T_{\rm EW}\simeq 160\,{\rm GeV}$ (as in the vast majority of baryogenesis models), then the early Universe would also feature an imbalance between Higgs doublets and anti-Higgs doublets\footnote{This happens because Higgs doublets decay into left handed fermions and anti-right handed ones. Left handed fermions feel the quantum anomalies which break $(B+L)_L$ and hence the chemical potentials of left and right handed fermions must be different, implying a non-zero Higgs chemical potential in the early Universe~\cite{Harvey:1990qw}.}. In the model of Ref.~\cite{Escudero:2025dei} a new inert Higgs doublet $H'$ is introduced precisely because,  having the same quantum numbers of the SM Higgs, it easily allows equilibrating interactions that transfer part of the Higgs doublet asymmetry 
into dark sector asymmetries. The presence of an inert Higgs doublet, which acts as a facilitator for the asymmetry transfer, is generic of 
most setups in this general class of asymmetric models. However, a dark sector constituted of only the $H'$ nevertheless does not allow a successful generation of a large enough dark matter $H'$ asymmetry (see \cite{Dhen:2015wra,Escudero:2025dei}). In this model, an additional complex singlet scalar, $\phi$, plays the role of the main asymmetric component of the dark matter and both states are stabilized under a new postulated discrete $Z_4$ symmetry, with charge $-2$ for $H'$ and $+1$ for $\phi$. With this symmetry choice and charge assignments, a key renormalizable portal that allows the $H$ asymmetry to partly flow into a dark matter asymmetry is\footnote{The $\lambda_{\phi 4} \phi^4$ and $\lambda_{\bar{\phi}^2HH'}\bar{\phi}^2 HH'$ interaction terms are perfectly allowed and also lead to particle transfer interactions. As shown in~\cite{Escudero:2025dei} as long as $\lambda_{\bar{\phi}^2HH'}$ is not within 10\% of $\lambda_{\phi^2 HH'}$ and $\lambda_{\phi 4}$ is much bigger than either $\lambda_{\bar{\phi}^2HH'}$ or $\lambda_{{\phi}^2HH'}$, the asymmetric dark matter mechanism is essentially left intact and hence the existence of these couplings does not pose any particular tuning on the scenario.}:
\begin{align}\label{eq:L_portal_1}
    \mathcal{L} \owns  -\frac{\lambda_{\phi^2 HH'}}{2} \phi^2 H^\dagger H' +{\rm h.c.}\,.
\end{align}
Via interactions of the type $\phi H \to \bar{\phi} H'$, this portal allows to generate an imbalance of $\bar{\phi}$ particles over $\phi$ ones that is directly tied to the baryon asymmetry of the Universe (because $\Delta H \propto \Delta B$). Another key asymmetry portal that is allowed by all the symmetries is the usual inert doublet quartic interaction:
\begin{align}
    \mathcal{L} \owns  -\frac{\lambda_{5}}{2} (H^\dagger H')^2 +{\rm h.c.}\,.
    \label{lambdaphiphiHHp}
\end{align}
This coupling serves two distinct roles. On the one hand, it tends to equilibrate the chemical potentials of $H'$ and $H$ and if that happens the $\phi-\bar{\phi}$ asymmetry is depleted. On the other hand, upon electroweak symmetry breaking, it provides a mass splitting between the two neutral components of the Higgs doublet:
\begin{align}
   \delta m \equiv  m_{A_0}-m_{H_0} \simeq \frac{\lambda_5}{2} \frac{v_H^2}{m_{H_0}} \,,
   \label{lambda5}
\end{align}
where $v_H \simeq 246\,{\rm GeV}$ is the Higgs vacuum expectation value and we have defined the inert Higgs doublet as $H' = (H^+, (H_0 + i A_0)/\sqrt{2} )^{T}$, where we have considered, without loss of generality, that $m_{A_0}>m_{H_0} $ (since this $H'$ does not couple to fermions these neutral scalars do not have a well defined CP parity).

The neutral components of the $H'$ doublet couple to the $Z$ boson off-diagonally.
Thus, a non-negligible $H'$ dark matter component is not excluded from a too fast $Z$-mediated dark matter to nucleon interactions in direct detection experiments, as the process can be made kinematically suppressed~\cite{Tucker-Smith:2001myb}. 
This requires the neutral mass splitting to be large enough, $\delta m \gtrsim 100\,{\rm keV}$.  Critically, the mass splitting cannot be too big either, otherwise the asymmetry of $\phi-\bar{\phi}$ particles will be zero and the mechanism for asymmetric dark matter will be inoperative, requiring $\delta m \lesssim 5\,{\rm MeV}$\footnote{The way a large enough $\phi$ asymmetry is created in this scenario is through asymmetric freeze-in of the asymmetry from partial transfer of the SM asymmetries. If the $\lambda_5$ interaction is large, then deep thermal equilibrium at a temperature of order $m_\phi$ will be established leading to the same asymmetries in $H'$ and $H$, $\Delta H' =\Delta H$. As the $\Delta \phi$ asymmetry is created from processes like $\phi H \to \bar{\phi} H'$, it is clear that, if $\mu_{H'} = \mu_{H}$, then $\mu_\phi \to 0$ and no asymmetry in $\phi$ could be generated. The successful transfer of asymmetries then restricts $\delta m \lesssim 5\,{\rm MeV}$.}.  This is different from the usual inelastic Higgsino dark matter models which do not involve any stringent upper limit on the neutral mass splitting. Thus, the neutral mass splitting is constrained to be in the following range,
\begin{align}
[\text{Direct Detection, see Fig.~\ref{fig:smallDelta}}]\,\,100\,{\rm keV}   \lesssim  \delta m \lesssim 5\,{\rm MeV} \,\,[\text{DM-asymmetry efficiency}]\,.
\label{lowersplitDD}
\end{align}

Beside the asymmetry portal interactions of Eqs.~\eqref{lambdaphiphiHHp} and~\eqref{lambda5}, the symmetry structure allows other terms in the scalar potential which do not transfer asymmetries:
\begin{align}
 V\owns m^2_{\phi}|\phi|^2+m^2_{H'}|H'|^2+\lambda_{\phi H} |\phi|^2 |H|^2+\lambda_{\phi H'} |\phi|^2|H'|^2+\lambda_3 |H|^2|H'|^2+\lambda_4|H^\dagger H'|^2\,.
\end{align}
A key one of this sort is the $\lambda_4$ coupling that leads to a tree-level mass splitting ($\Delta M|_{\lambda_4}$) between the charged and neutral components of the multiplet:
\begin{align}
    \Delta M \equiv m_{H^\pm}-m_{H_0} = \Delta M|_{\lambda_4} +\Delta M|_{\rm EW}  \simeq -\frac{\lambda_4}{4} \frac{v_H^2}{m_{H_0}}+\Delta M|_{\rm EW}\,,
    \label{lowerchargedsplit}
\end{align}
where $\Delta M|_{\rm EW}$ is generated by electroweak corrections~\cite{Cirelli:2005uq} and, as depicted in Fig.~\ref{fig:bigDelta_EWcorr_Money}, it is $\Delta M|_{\rm EW}\lesssim 300\,{\rm MeV}$. Interestingly, this $\lambda_4$ interaction also contributes to direct detection scatterings which are rather constrained, $\sigma_{\rm SI} \propto |\lambda_3+\lambda_4+\lambda_5|^2$. This means that, in the absence of cancellations, current direct detection limits imply an upper bound on $\lambda_4$, which written in terms of mass splitting reads as $\Delta M \lesssim 1\,{\rm GeV}$, apart from a narrow window for $m_{H'}\lesssim 60\,{\rm GeV}$, see Fig.~\ref{fig:bigDelta_EWcorr_Money}. 
Thus, together with Eq.~(\ref{lowersplitDD}) one concludes that the inert Higgs doublet must generically be compressed.

\begin{figure}[t]
    \centering
    \includegraphics[width=0.6\linewidth]{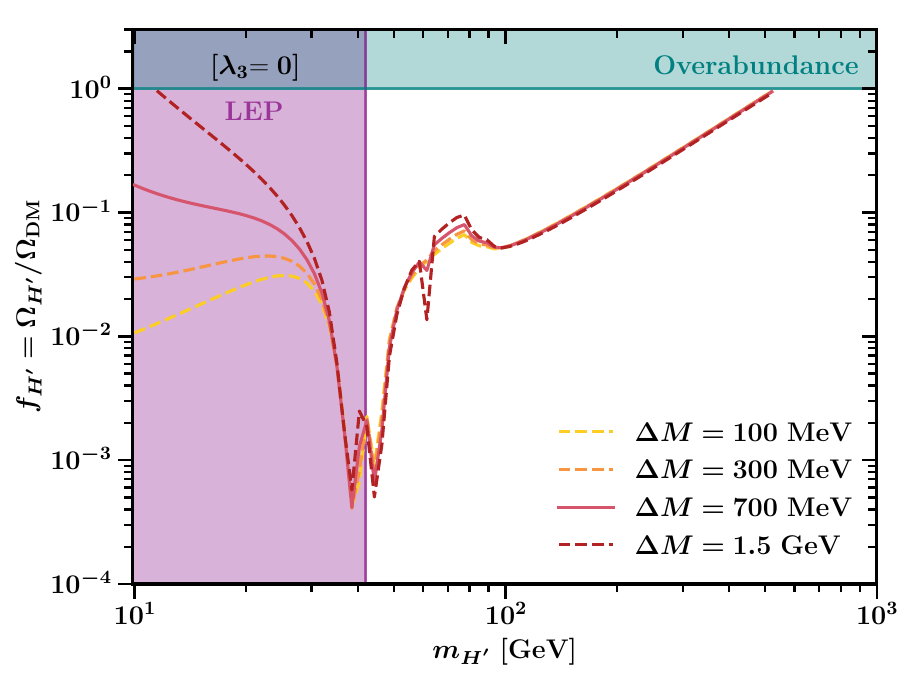}
    \vspace{-0.3cm}
    \caption{Relic abundance for the $H'$ dark matter component resulting from thermal freeze-out as calculated using the \texttt{micrOMEGAs} code~\cite{micrOmega,Alguero:2023zol,micrOMEGAs630} -- the rest of the dark matter in the scenario is made of the asymmetric $\phi$ states with a number density directly related to the baryon asymmetry of the Universe. We show the fraction of the dark matter energy density of the Universe in the form of $H'$ particles for four different choices of the charged-neutral mass splitting, $\Delta M$. We can clearly see that there is no relevant difference between them in the region which is not excluded by LEP~\cite{Blinov:2015qva}. Given this weak dependence, in what follows, we effectively use the $\Delta M=700$ MeV results for any value of the $\Delta M$ parameter. Note that this figure was made using $\delta m = 5$ MeV and $\lambda_3 = 0$.}
    \label{fig:relicHp}
\end{figure}

The mass range for $H'$ is constrained by two requirements. On the one hand, $m_{H'}\lesssim 580\,{\rm GeV}$ so that its thermal abundance does not exceed the observed dark matter density~\cite{Bottaro:2022one,Cirelli:2005uq}. On the other hand, and quite interestingly, a compressed inert Higgs doublet mass spectrum turns out not to have been constrained by LEP-2~\cite{Blinov:2015qva}. In fact, the only absolute limit on its mass arises from the requirement that it must not contribute significantly to the invisible decay width of the $Z$ boson, which implies $m_{H'} \gtrsim m_Z/2 \simeq 46\,{\rm GeV}$. This tells us that the key window of interest for this particle is:
\begin{align}
[Z\text{-invisibles}]\,\,46\,{\rm GeV} \lesssim     m_{H'}\lesssim 580\,{\rm GeV} \,[\text{DM overproduction}]\,.
\end{align}
A key ingredient for its phenomenology is that the $H'$ relic density can be fully predicted as a function of its mass (since it is set by SM interactions), as given in Fig.~\ref{fig:relicHp} (obtained using the \texttt{micrOMEGAs} code~\cite{micrOmega,Alguero:2023zol,micrOMEGAs630}). These results have a very weak dependence on the  mass splitting, $\Delta M$. Critically, for the mechanism, the smaller the density of $H'$ is in the Universe, the more asymmetric dark matter is, as the rest of the dark matter density will be in the form of $\phi$ particles. 
This allows to set predictive upper bounds on the $H'$ interactions to the $Z$ and to the $H$ bosons, in particular from direct detection experiments\footnote{We note that given the results from \texttt{micrOMEGAs} the $H'$ saturates the dark matter relic density for $m_{H'}\simeq 538\,{\rm GeV}$ which is slightly lower than the refined value reported in~\cite{Bottaro:2022one} of $m_{H'} = 580\pm 10\,{\rm GeV}$ which includes Sommerfeld enhancements and bound state formation in the relic abundance calculation. These effects become increasingly smaller the lighter the $H'$ is and we disregard the uncertainty from not considering them in our calculation.}.

Next, concerning the dominant dark matter component, i.e., the $\phi$ asymmetric component, the scenario implies that $m_\phi$ must lie within the following range 
\begin{align}
[\text{Asymmetry efficiency}]\,\,700\,{\rm GeV}\lesssim     m_\phi \lesssim 30\,{\rm TeV}\,\,[\text{perturbativity}].
\end{align}
Beyond the fact that the scenario requires $\phi$ to be heavier than $H'$, the origin of the lower bound is two-fold. First, only a fraction of the SM asymmetries is transferred to the dark matter sector, so that the dark matter particle must be well above the proton mass. Second, the asymmetry must be created sufficiently before the electroweak phase transition since, after it, there are no more $H$ and $H'$ asymmetries that could source any $\phi$ asymmetry. 
As for the upper limit, it comes from the fact that, in this scenario, the $\phi \bar{\phi} \to H'\bar{H}'$ interactions, which need to be efficient enough to annihilate particles and antiparticles, would turn non-perturbative. As we will discuss too, this dominant $\phi$ component can be probed by direct detection experiments as well, thus offering another potential complementary signature of the scenario.

In summary, the scenario we consider contains two dark matter particles $\phi$ and $H'$. $\phi$ is the main dark matter component and its density is directly linked to the baryon asymmetry of the Universe. $H'$ facilitates the asymmetry transfer from the baryons to the dark matter and it turns out to be a sub-dominant fraction of the dark matter, see Fig.~\ref{fig:relicHp}. Quite remarkably, and in contrast with naive expectations, $H'$ can be as light as $46\,{\rm GeV}$, despite being electroweakly charged. In the next sections, we focus on investigating the direct detection signatures of this inert Higgs doublet, as well possible searches for it at the LHC.

\section{Direct Detection}\label{sec:DD}

\begin{figure}[t]
    \centering
    \includegraphics[width=0.7\linewidth]{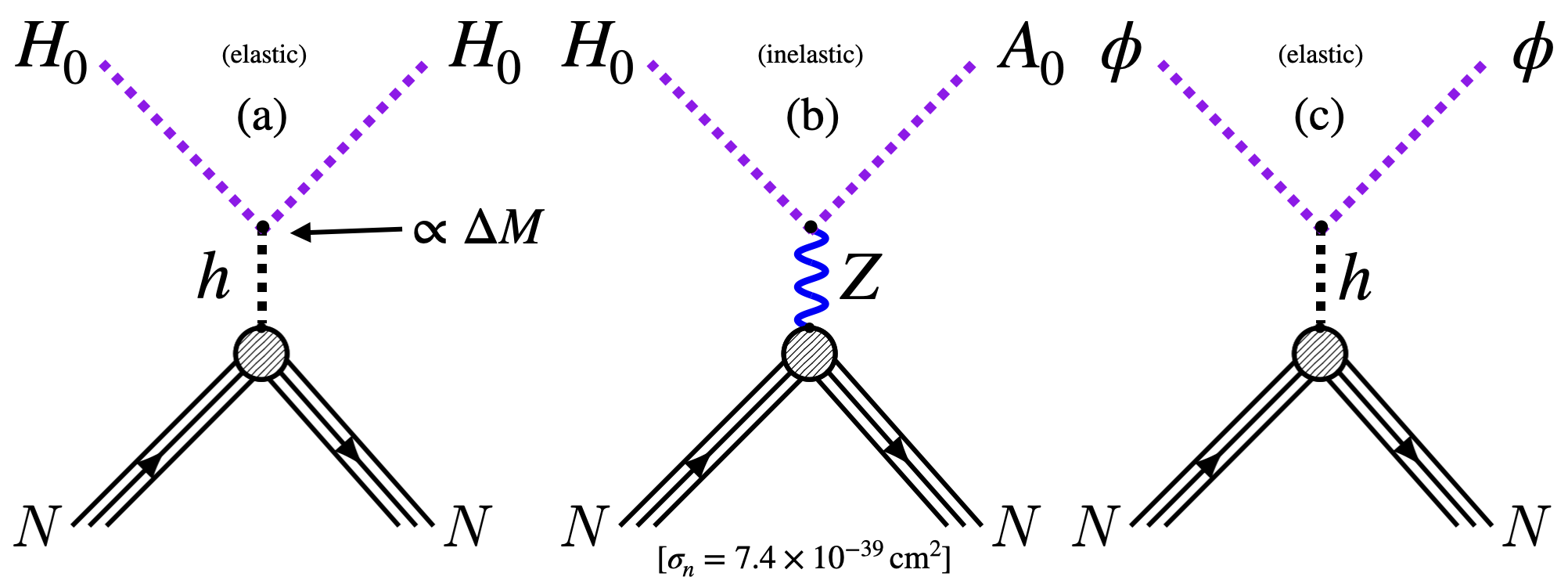}
    \caption{The three different channels for direct detection of dark matter within the model of study. $H_0$ is a sub-dominant component of dark matter (see Fig.~\ref{fig:relicHp}), but can interact both elastically as well as inelastically with nuclei. In the absence of cancellations, the elastic interaction rate is directly proportional to the induced charged-neutral mass splitting, $\Delta M|_{\lambda_4}$, which is a key parameter for collider searches as it controls the lifetime of $H^\pm$. $H_0$ also has an inelastic scattering channel mediated by the $Z$ boson and whose cross section is therefore fully predicted and its rate depends only upon $\delta m$ and $m_{H'}$. Finally, $\phi$, the main component of dark matter, can also interact elastically with nuclei as mediated by the Higgs boson. The rates are written in Eq.~\eqref{eq:sigman}.}
    \label{fig:DDcartoon}
\end{figure}

Our scenario features three different types of interaction channels that can be searched for at dark matter direct detection experiments, as illustrated in Fig.~\ref{fig:DDcartoon}. The most interesting ones are those arising from the $H_0$ state as, in the absence of strong cancellations, they give information about the mass difference between the charged and neutral components of the doublet (see Eq.~\eqref{eq:H0elSigma}). The inelastic channel (b), mediated by the $Z$ boson, is also quite interesting because it has a rate that is fully predicted by the model as a function of the neutral mass splitting value, $\delta m$. With these the three channels depicted in Fig.~\ref{fig:DDcartoon} in mind, we investigate the region of parameter space which is allowed or favored by current results from direct detection dark matter experiments, and in particular from LZ~\cite{LZ:2024zvo,LZ:2026axp}. Critically, we will also investigate the prospects from the next generation of experiments in the upcoming years.

\subsection{Dark Matter Interactions}
\label{sec:DD_interaction}

A generic dark matter particle, $\chi$, that scatters via spin independent interactions with nuclei features a differential cross section with a nucleus, $N$, given by
\begin{equation}
    \frac{d\sigma}{dE_R} = \frac{1}{2v^2}\frac{m_N \sigma_{n\chi}}{\mu_{n\chi}^2}\frac{(f_p Z + f_n (A-Z))^2}{f_n^2} F^2(E_R)\,,
\end{equation}  
with $v$ the incoming dark matter velocity, $m_\chi$ the dark matter mass, $m_N$ the mass of the nucleus, $m_n$ the mass of a nucleon, $\sigma_{n\chi}$ the dark matter-neutron cross section at zero momentum transfer and $\mu_{n\chi}$ the reduced mass of the $\chi\!-\!n$ system. $A$ and $Z$ are, respectively, the mass and atomic numbers of the nucleus,  $f_p$ and $f_n$ are parameters which control the relative strength of the interactions with neutrons and protons\footnote{For channels (a) and (c) $f_n = f_p$ as they are mediated by the Higgs boson. On the other hand, for channel (b) one has $f_n \simeq 1$ and $f_p \simeq - 0.075$ as the $Z$ coupling to protons is accidentally suppressed by $4s_W^2-1 \simeq- 0.075$, which is a small number since $s_W^2 \simeq 0.231$.}, and $E_R$ is the recoil energy of the outgoing nuclei in the collision which is the key observable. Finally, $F(E_R)$ is the nuclear form factor which for xenon isotopes we take from~\cite{Vietze:2014vsa} and for other nuclei we consider the Lewin-Smith parametrisation~\cite{Lewin:1995rx} of the Helm form factor, as in~\cite{McCabe:2010zh}.

The number of events from dark matter-nucleus scattering is given by (see e.g.~\cite{Cerdeno:2010jj,McCabe:2010zh}):
\begin{equation}
    \frac{dN}{dE_R} = \frac{\epsilon\rho_\chi}{m_\chi m_N}\int_{v_{\rm min}}^{\infty} v f_{\oplus}(\vec v, \vec v_e)\frac{d\sigma}{dE_R}d^3v \,,
\end{equation}
where $N$ is the number of events, $\epsilon$ is the exposure of the experiment, and $f_{\oplus}$ is the dark matter velocity distribution on Earth. The scattering rates for elastic nuclear recoils are fairly insensitive to the local dark matter distribution and for those we consider $f_{\oplus}$ as being described by the Standard Halo Model (SHM) with a dark matter local density of $\rho_\chi=0.3\, {\rm GeV/cm^3}$, $v_e = 253.67$~km/s for the Earth's velocity and a escape velocity of $v_\mathrm{esc}$ = 544 km/s. The rate for inelastic scatterings can depend significantly more on the actual shape and tails of the $f_{\oplus}$ distribution. In this case,  we consider both the Standard Halo Model and the possibility of a small population of high velocity particles in the tail of the distribution arising from dark matter particles from the past interaction between the Large Magellanic Cloud (LMC) and the Milky Way (MW). In particular, and as summarized in Table~\ref{tab:DM_distributions}, we consider three cases for the relevant velocity integral as displayed in the top panel of Fig.~\ref{fig:smallDelta}: 1) SHM, as before, 2) MW-LMC, as from the results of the simulation reported in Figure 2 of~\cite{Reynoso-Cordova:2024xqz}, see also~\cite{Smith-Orlik:2023kyl,Bozorgnia:2026emp}, and 3) MW-LMC-extrahigh which depicts a dark matter distribution found to allow a good fit from $H'$ interactions to the high energy recoil observed by LZ, while allowing the model to be consistent with the strong bounds from neutrino emission from the sun, 
see below.\footnote{This dark matter distribution is not excluded on observational grounds but it is neither predicted in simulations so far, see~\cite{OHare:2026nqi,Smith-Orlik:2023kyl,Folsom:2025lly}. It just shows what would it take it for the dark matter distribution to be in order to fit the signal within our model while being compatible with other observations.}.

\begin{figure}[t!]
    \centering
    \includegraphics[width=0.95\linewidth]{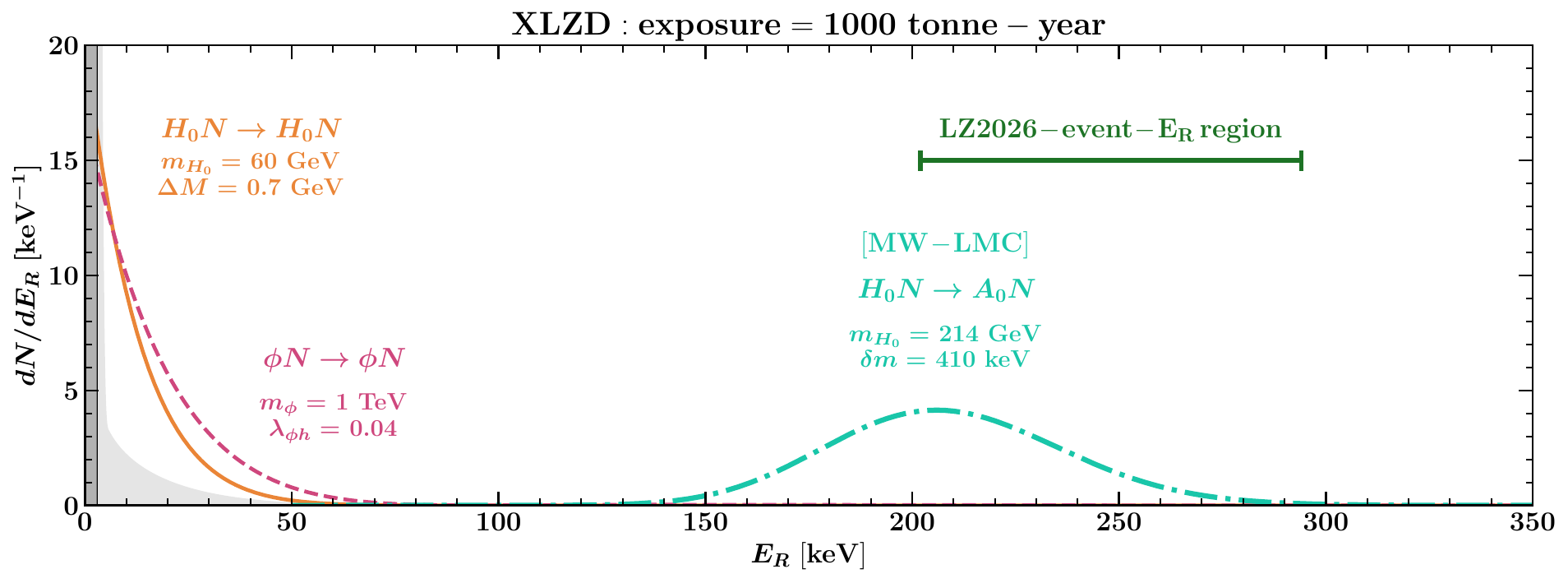}
    \vspace{-0.3cm}
    \caption{Expected recoil energy spectra at the XLZD-1000 experiment for the elastic $H_0$ and $\phi$ channels, as well as the inelastic $H_0$ channel for the indicated benchmark values of the parameters. The case with $\delta m = 410\,{\rm keV}$ is only observable when a high energy tail in the dark matter distribution is considered, such as the case of the MW-LMC dark matter case, used here. The light gray region in the lower left part corresponds to the neutrino floor energy spectrum, and the green band region spanning $200-300\,{\rm keV}$ corresponds to the kinematical region where the event was observed by LZ at $2\sigma$ confidence~\cite{LZ:2026axp}.}
    \label{fig:dNdE}
\end{figure}

To get the number of events, the differential formula needs to be integrated and one needs to take into account the experimental efficiency $\varepsilon(E_{\rm R})$ as well as the fact that experiments have a finite energy resolution $\Delta E_R$. We then calculate the number of events between a range of reconstructed energies $E_{\rm reco}^{\rm min}$ and $E_{\rm reco}^{\rm max}$ as 
\begin{subequations}
\begin{align}
    N &=  \int_{E_{\rm reco}^{\rm min}}^{E_{\rm reco}^{\rm max}} dE _{\mathrm{reco}} \int_{0}^{\infty} dE_\mathrm{R} \frac{e^{-\frac{(E_{\rm reco}-E_{\rm R})^2}{2\Delta E_R^2}}}{\sqrt{2\pi}\Delta E_R}\frac{dN}{dE_\mathrm{R}}\varepsilon(E_\mathrm{R})~,\\
    &= \int_0^{\infty} d E_R \frac{dN}{dE_R} \varepsilon(E_R) P_{[E_{\rm reco}^{\rm min},E_{\rm reco}^{\rm max}]}(E_R)\,\quad \text{with} \quad  P_{E_a,E_b}(E_R) = \frac{1}{2} \left({\rm Erf}\left[\frac{E_b-E_R}{\sqrt{2}\Delta E_R}\right]-{\rm Erf}\left[\frac{E_a-E_R}{\sqrt{2}\Delta E_R}\right]\right)\,,
\end{align}
\end{subequations}
where $\Delta E_R(E_R)$ is the energy resolution which typically depends upon $E_R$, and ${\rm Erf}$ is the error function that appears upon explicit integration of the Gaussian kernel. It is important to highlight that there is a minimum velocity for which a nuclear recoil of energy $E_R$ can be obtained and reads:
\begin{align}\label{eq:vmin_kine}
    v_{\rm min} = \frac{1}{\sqrt{2m_N E_R}} \left(\frac{m_N E_R}{\mu_{\chi N}}  + \delta m \right)\,,
\end{align}
where $\mu_{\chi N}$ is the reduced mass of the dark matter-nuclei system and $\delta m $ is the possible mass splitting between the outgoing and incoming dark matter particles. 

\textbf{Scattering Interaction Rates:} The scattering cross sections for the three different channels with a neutron/proton read:
\begin{subequations}\label{eq:sigman}
    \begin{align}\label{eq:H0elSigma}
    \sigma_n &= |\lambda_3+\lambda_4+\lambda_5|^2\frac{f_N^2}{4\pi}\left(\frac{\mu_{H_0 n}m_n}{m_{H_0}m_h^2}\right)^2 \,\,  [f_n = f_p = 1] \ \quad \quad\quad\,\,\,\, \,\,\text{$H_0$ + N $\rightarrow$ $H_0$ + N (elastic)} \\
  \sigma_n &= 10^{-47}\,{\rm cm}^2\,\left(\frac{|\Delta M|_{\lambda_4}|}{500\,{\rm MeV}}\right)^2 \quad \,\,\,\,\quad \quad [\lambda_3 = \lambda_5 = 0] \nonumber    \\ \label{eq:H0inelSigma}
     \sigma_n &= \frac{1}{2\pi} G_F^2 \mu_{H_0 n}^2 \simeq 7.4\times 10^{-39}\,{\rm cm^2} \,\,\,\,\,\,\,\,[f_n = 1, f_p\simeq -0.075] \quad \text{$H_0$ + N $\rightarrow$ $A_0$ + N (inelastic)} \\
     \sigma_n &= |\lambda_{\phi H}|^2\frac{f_N^2}{4\pi}\left(\frac{\mu_{\phi n}m_n}{m_{\phi}m_h^2}\right)^2 \,\,\,\,\qquad\qquad[f_n  = f_p= 1] \qquad \quad \,\,\,\,\,\,\,\,\,\,\,\,\,\text{$\phi$ + N $\rightarrow$ $\phi$ + N (elastic)} \,.
    \end{align}
\end{subequations}
In the second line we have assumed $\lambda_3 =  \lambda_5 = 0$ and $\Delta M|_{\lambda_4}$ is the mass splitting evaluated at tree level (see Eq. \ref{lowerchargedsplit}). Note that if $H'$ is to allow a successful transfer of asymmetries to $\phi$ in the early Universe then $\lambda_5$ has a negligible contribution in the first expression. $f_N \simeq 0.308$ is the effective Higgs-nucleon coupling~\cite{Hoferichter:2017olk}.

Given these interactions and the formulae above, we show in Fig.~\ref{fig:dNdE} a combination of potential signals in a xenon experiment with $1000$ tonne-year of exposure as being representative of what may be observed at an experiment like XLZD~\cite{XLZD:2024nsu}. We show the expected signal from the three different channels together with the irreducible neutrino background that we take from the \texttt{NeutrinoFog} code of~\cite{OHare:2021utq}, see also~\cite{Billard:2013qya}. These spectra correspond to benchmarks which are currently allowed and, due to their different spectral shape, may be disentangled.

\begin{center}
\small
\textbf{Current Constraints and Expected Sensitivity from Direct Detection Dark Matter Experiments}
\end{center}

Our goal in this section is to understand what is the parameter space in the model that is currently allowed and to see what regions of parameter space can be tested with the next generation of experiments. In terms of current constraints, we focus on the LZ experiment for both the elastic and inelastic channels, see~\cite{LZ:2024zvo,LZ:2026axp} but also~\cite{XENON:2025vwd,PandaX:2024qfu} for comparable results. With regards to the future sensitivity, we consider various experiments using different types of isotopes. Specifically, PandaX-xT~\cite{PANDA-X:2024dlo} and XLZD~\cite{XLZD:2024nsu} for xenon, ARGO~\cite{GalbiatiARGO} for argon and  RES-NOVA~\cite{Res-Nova,Alloni:2026xdf} for heavy lead based experiments. In the context of elastic scatterings, we effectively take the results derived from each experimental collaboration, while for inelastic ones we craft our own likelihoods as discussed below.

\subsection{Elastic Scatterings: Implications for the charged-neutral mass splitting} 

\begin{figure}[t]
    \centering
    \begin{tabular}{cc}
       \includegraphics[width=0.47\linewidth]{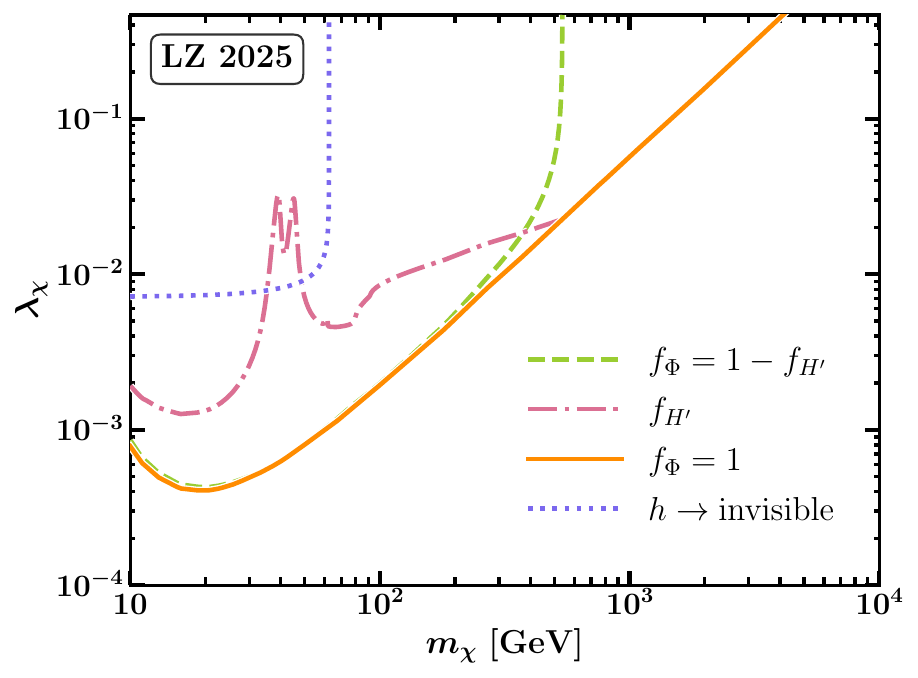}    &    \includegraphics[width=0.44\linewidth]{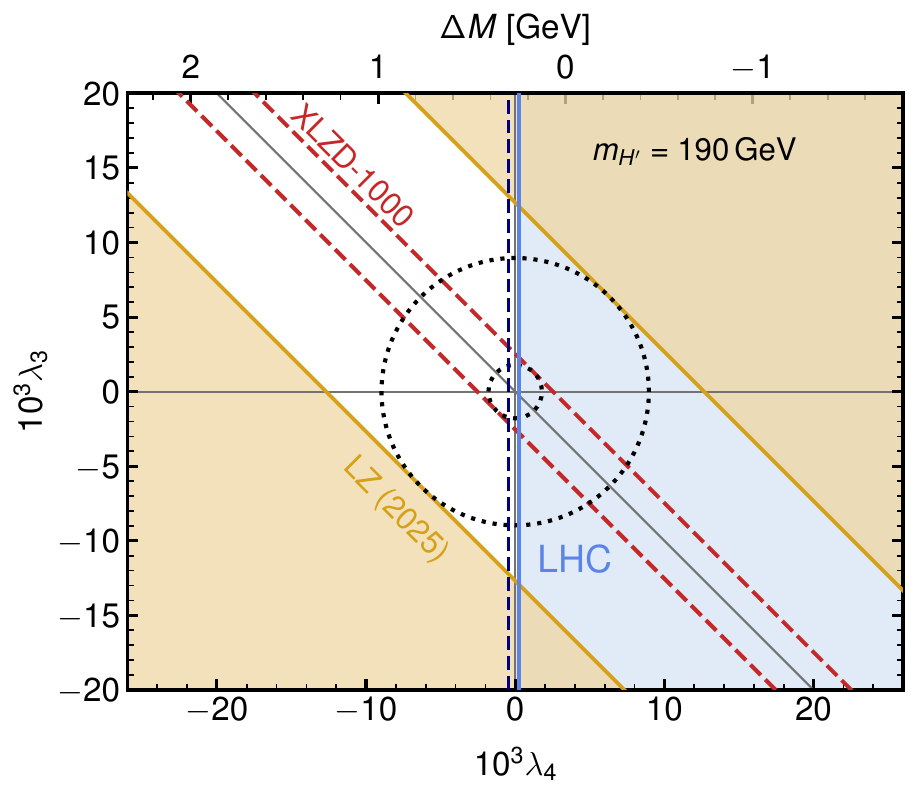} \\
        \includegraphics[width=0.47\linewidth]{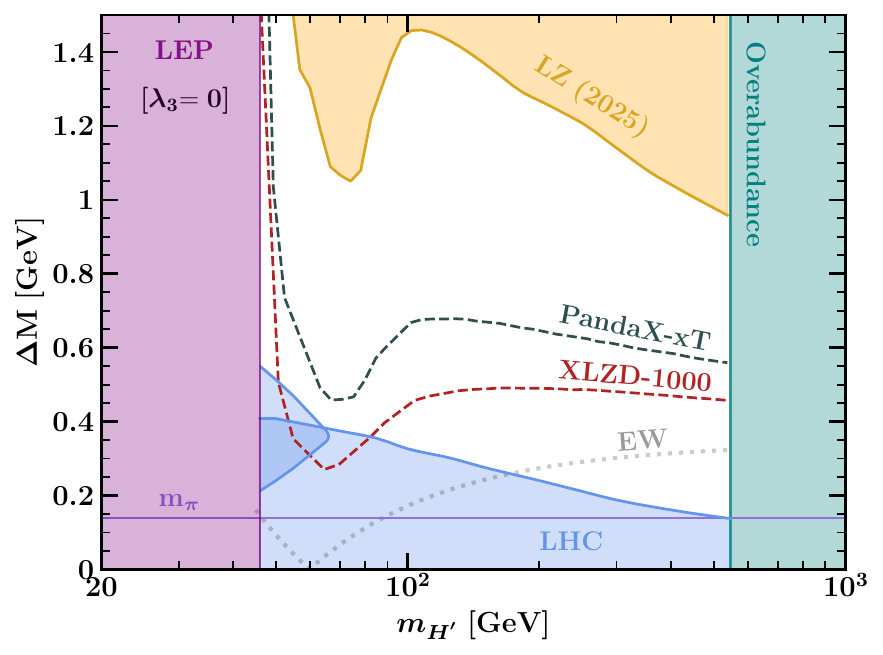}  &    \includegraphics[width=0.47\linewidth]{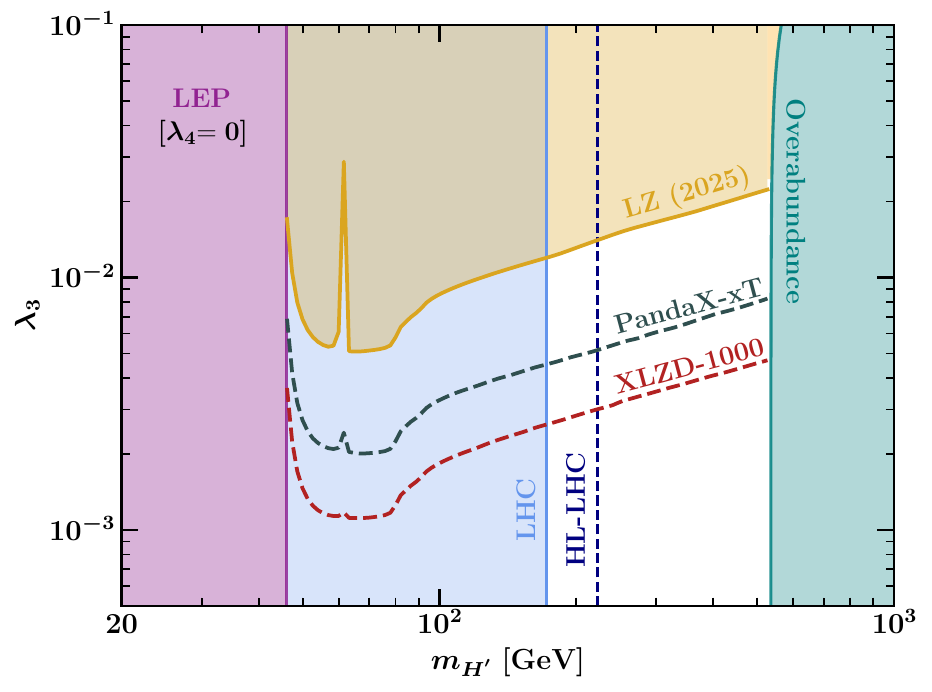}
    \end{tabular}
    \vspace{-0.3cm}
    \caption{Summary of the results in terms of current constraints and future sensitivities from elastic interactions in the model under consideration with $H'$ and $\phi$ as dark matter components. \textit{Upper left panel:} Bounds on the quartic couplings between these scalars and the Higgs boson. We show the current limits from LZ as well as the one coming from the invisible width of the Higgs, see Section~\ref{sec:collider}. 
    The pink and green lines give the direct detection upper bound on $|\lambda_3+\lambda_4|$ and $\lambda_{\phi H}$ respectively, as a function of $m_{H'}$. The orange line gives the direct detection upper bound on $\lambda_{\phi H}$ as a function of $m_\phi$, assuming that the $H'$ relic density is small, i.e. for $m_{H'}$ well below 580 GeV. \textit{Upper right panel:} Parameter space allowed in the $\lambda_3-\lambda_4$ plane for the case $m_{H'} = 190\,{\rm GeV}$. One sees that unless there are strong cancellations (i.e. well outside the circle) direct detection requires the doublet to be very degenerate. \textit{Lower left panel:} bounds on $\Delta M$ assuming $\lambda_3 = 0$. We clearly see that direct detection experiments are key to probe this mass difference. \textit{Lower right panel:} limits on $\lambda_3$ assuming $\lambda_4 =0 $ where we again see the complementarity between the various experiments. }
    \label{fig:elastic_results}
\end{figure}

\begin{figure}[t!]
    \centering
    \includegraphics[width=0.9\linewidth]{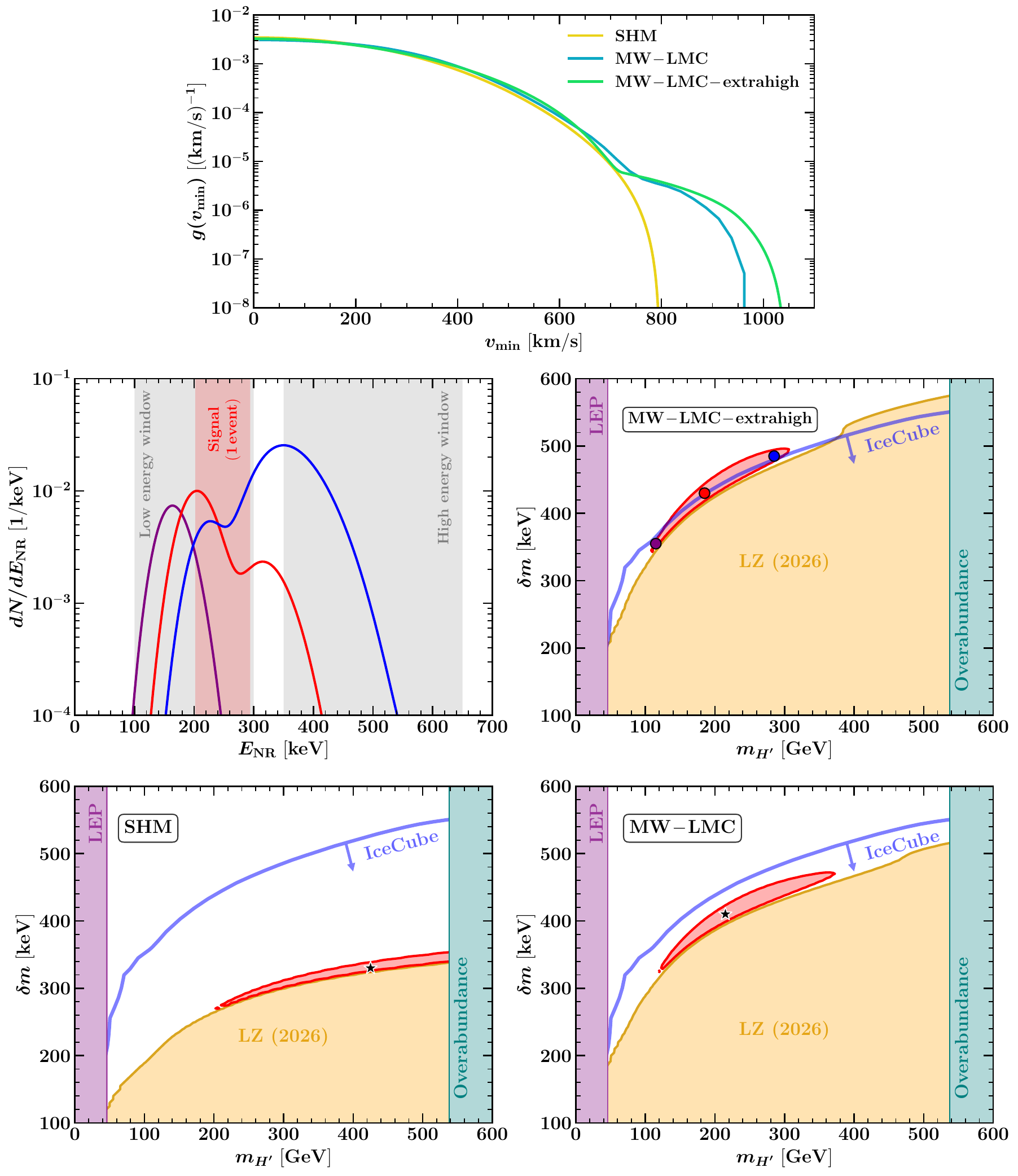}
    \caption{Inelastic dark matter phenomenology for a light inert Higgs doublet. Yellow bands indicate current constraints from LZ ($\Delta \chi^2 = 9.21$, 99\% CL), red regions represent the favored ($\Delta \chi^2 = 4.61$, 90\% CL) regions as relevant for explaining the high energy recoil LZ event, and the blue regions correspond to bounds on this particle from neutrinos arising from dark matter annihilations in the sun as constrained by IceCube (90\% CL).  \textit{Top panel:} Local dark matter distributions considered. \textit{Middle left panel:} Smeared energy recoil spectra as relevant for the LZ~\cite{LZ:2026axp} search. We show the various regions considered in our likelihood including the high energy sidebands. \textit{Middle right panel:} Results considering the MW-LMC-extrahigh distribution. \textit{Bottom left panel:} Results considering the SHM. \textit{Bottom right panel:} Results considering the MW-LMC model. We see that only with a dark matter distribution featuring a high velocity tail ($v_{\rm min} \gtrsim 1000\,{\rm km/s}$) could this model fit the signal while being in agreement with the lack of high energy neutrino emission from the sun.}
    \label{fig:smallDelta}
\end{figure}

Fig.~\ref{fig:elastic_results} displays a summary of the current limits on the inert Higgs doublet derived from the null searches for elastic dark matter scatterings with nuclei. In the upper left panel, we show the current limits on the couplings $|\lambda_{3}+\lambda_4+\lambda_5|$ and $\lambda_{\phi H}$ as relevant for the $H'$ and $\phi$ dark matter candidates, respectively. In the asymmetric dark matter model under consideration, $\lambda_5\lesssim 10^{-5}$ and hence too small to lead to an elastic scattering signal. In the upper right panel, we consider the case $m_{H'} = 190\,{\rm GeV}$ as an example of how the bounds from direct detection constrain the model. One can see that in the absence of strong cancellations, i.e. $\lambda_3 \neq -\lambda_4$, the bound applies roughly individually to each coupling. Since $\lambda_4$ controls the charged-neutral mass splitting, this in turn means that the inert Higgs doublet should be degenerate $\Delta M \lesssim 1.2\,{\rm GeV}$ given current constraints.

The lower left panel of Fig.~\ref{fig:elastic_results} shows the results in the $\Delta M$-$m_{H'}$ plane under the assumption that $\lambda_3 = 0$. PandaX-xT and XLZD are aiming to improve the current sensitivity by factors $10-25$ on elastic scattering interactions and, as can be seen in the figure, this has an enormous impact in the parameter space that will be tested. In particular, while currently $\Delta M \lesssim 1.2\,{\rm GeV}$, the next generation of experiments will test down to $\Delta M \lesssim (0.3-0.5)\,{\rm GeV}$ in a very complementary way to the LHC (see Section~\ref{sec:collider}). The lower right panel instead shows the case $\lambda_4 = 0$ and hence with $\Delta M = \Delta M|_{\rm EW}$. We see that the next generation of dark matter experiments will be very important to test regions of this parameter space as well. Overall, we learn that the next generation of direct detection experiments combined with the results from HL-LHC are expected to broadly test the model.

In Fig.~\ref{fig:elastic_results} we show the 90\% sensitivity from these dark matter experiments and one wonders about whether in the presence of a future dark matter signal for elastic scatterings, parameter reconstruction of the particle that generated it could be possible. As it is well-known, for high dark matter mass this is very difficult to achieve due to a strong degeneracy between the interaction cross section and the mass of the dark matter, ${\rm rate} \propto \sigma_{\chi n}/m_\chi$. Nevertheless, as we show in Appendix~\ref{app:DD}, if the inert Higgs is lighter than $\sim 75\,{\rm GeV}$ and its interaction rate is only a factor $\sim 3$ smaller than the current limit excluded by LZ, its mass and interaction rate may still be reconstructed with $\sim 2\sigma$ confidence. We note that in this context, combining high exposure xenon experiments with an argon one like ARGO would help in narrowing down this parameter space.

\begin{table}[t]
    \centering
    \small
    \resizebox{\textwidth}{!}{%
    \begin{tabular}{l|c}
        \hline \hline
        Dark Matter Distribution $\,\,\,$ & Description \\
        \hline \hline
        SHM &
        \parbox[t]{0.8\textwidth}{
        Standard Halo Model as described in the text. Following Ref.~\cite{Baxter:2021pqo}.
        } \\
\hline
        MW-LMC &
        \parbox[t]{0.8\textwidth}{
        Dark matter distribution obtained from a simulation of a Milky Way analogue with a satellite resembling the mass and present-day orbit of the LMC. From Ref.~\cite{Reynoso-Cordova:2024xqz}.
        } \\
\hline
        MW-LMC-extrahigh &
        \parbox[t]{0.8\textwidth}{
        An ad-hoc distribution constructed in this work to investigate how high does the high-velocity tail of the dark matter distribution on Earth needs to be so that a light inert Higgs doublet could explain the LZ event while remaining consistent with solar capture limits.
        } \\
        \hline\hline
    \end{tabular}}
    \vspace{-0.35cm}
    \caption{Local dark matter velocity distributions considered in this work. See the top panel of Fig.~\ref{fig:smallDelta} for the shape of the velocity integral as a function of $v_{\rm min}$.}
    \label{tab:DM_distributions}
\end{table}

\begin{table}[t]
    \centering
    \resizebox{0.7\textwidth}{!}{%
    \begin{tabular}{c|c|ccccc}
        \hline
        \hline
        Dataset
        & LZ 2024~\cite{LZ:2024zvo}
        & \multicolumn{5}{c}{LZ 2026~\cite{LZ:2026axp}} \\\hline \hline
        $E_{\rm R}$ [keV]
        & $1$--$75$
        & $100$--$200$
        & $200$--$300$
        & $350$--$450$
        & $450$--$550$
        & $550$--$650$ \\
        $N_{\rm obs}$
        & 8 & 0 & 1 & 0 & 0 & 0 \\
        $N_{\rm exp}$
        & 12 & 0 & 0.01 & 0 & 0 & 0 \\
        Efficiency
        & LZ-2024
        & LZ-2026
        & LZ-2026
        & $\epsilon_{\rm SB}$
        & $\epsilon_{\rm SB}$
        & $\epsilon_{\rm SB}$ \\
        Exposure [tonne-year]
        & 4.2
        & \multicolumn{5}{c}{2.84} \\
        \hline\hline
    \end{tabular}}
    \vspace{-0.2cm}
    \caption{Energy bins, events, exposure and efficiency for our LZ likelihood as relevant for the inelastic scattering channel. We nominally take $\epsilon_{\rm SB}=0.96$ for the high-energy side-band, assuming the same typical efficiency as in the $100$--$300\,{\rm keV}$ region. We smear the recoil spectrum with a Gaussian of $\Delta E_{\rm R} = 1.46\,{\rm keV} \sqrt{E_{\rm R}/{\rm keV}}$~\cite{McCabe:2026crm}. We conservatively take $N_{\rm exp} = 0.01$ in the signal bin as a proxy for the background expectation in the whole energy window, as reported in the caption of Fig. 5 of~\cite{LZ:2026axp}. }
    \label{tab:LZ_likelihood}
\end{table}

\subsection{Inelastic Scatterings: Can the model fit the high energy nuclear recoil event observed at LZ?} \label{sec:DD_inelastic}

Another key parameter in the model is the mass splitting between the two neutral doublet components. As discussed in Section~\ref{sec:model}, the successful transfer of the baryon asymmetry into a dark matter asymmetry requires it to be $\delta m \lesssim 5\,{\rm MeV}$. Since the mass splitting controls the inelastic scattering between $H_0 N \to A_0 N$, it can be constrained by direct detection experiments. 

The most powerful search to constrain the $\delta m$ parameter has been recently performed by the LZ experiment~\cite{LZ:2026axp}. Interestingly, the collaboration has reported a high energy nuclear recoil event at a reconstructed energy of $E_{R} = 248\pm 23\pm 23\,{\rm keV}$ with the probabilities of known backgrounds to produce one such event with its observed ionization and scintillation yields being very low. This event has lead to ample discussion in the literature as it may be interpreted as a dark matter signal. To understand this interpretation in the context of a light inert Higgs doublet dark matter we build a simple version of the LZ likelihood using a product of Poisson distributions as displayed in Table~\ref{tab:LZ_likelihood}. We have a similar set up as in~\cite{Asadi:2026iot} and consider bins of $100\,{\rm keV}$, with one observed event in the 200-300 keV bin, and also considering the high energy sidebands, as highlighted in~\cite{Rodd:2026tyn}. The observed event was seen in June, precisely when the dark matter wind is expected to be stronger, but we consider for simplicity no modulation in the dark matter signal, knowing that this is nevertheless a key aspect to test its potential dark matter origin~\cite{McCabe:2026crm}. 

Fig.~\ref{fig:smallDelta} shows the current constraints on this inelastic mass splitting as a function of the $H'$ mass, for the three choices of the local dark matter velocity distribution described above and summarized in Table~\ref{tab:DM_distributions}. We see that LZ leads to rather tight limits on the $\delta m$ parameter and also that there are regions of parameter space which can in principle fit well the signal. For the models that contain a high velocity contribution to the dark matter distribution we readily see that there are closed contours. As depicted in the middle left panel showing the smeared differential rate, this has to do with the lack of observed events at the high energy sidebands. Reducing the WIMP detection efficiency in this kinematical region would allow these red contours to extend to higher masses. We have explored the reach for the lead based RES-NOVA experiment and concluded that with 0.17 tonne-year of exposure is unlikely to have sensitivity to the signal region even for the SHM case, we refer to Appendix~\ref{app:DD} for more details. 

Very soon after the LZ event was announced and interpreted as a possible signal of electroweakly interacting dark matter see e.g.~\cite{Fan:2026kxx,Langhoff:2026ujr}, it was pointed out in Ref.~\cite{Pospelov:2026ewn} that these types of particles (like the Higgsino or the inert doublet) would suffer from very strong constraints from the lack of observed high energy neutrino events from the sun. The idea is that dark matter will be captured in the sun, thermalize and annihilate into high energy neutrinos which can be detected at experiments such as Super-Kamiokande or IceCube, see~\cite{IceCube:2025fcu}. We have implemented the original analysis of~\cite{Pospelov:2026ewn}, see~\cite{Nussinov:2009ft,Menon:2009qj} for previous references, and~\cite{Bose:2026ndd,DiMauro:2026dqp,Nguyen:2026lui,Luengas:2026tlf} for recent ones, as well as Appendix~\ref{sec:solar_appendix}, where we give all relevant technical details of our analysis. Fig.~\ref{fig:smallDelta} depicts in blue the limits from the lack of neutrino signals in IceCube from the sun as recasted for our inert Higgs doublet model, considering in particular it is a sub-dominant dark matter component. The limits from the sun are effectively independent of the high velocity tail of the distribution and hence common for all panels.

From Fig.~\ref{fig:smallDelta}, we see that the lack of observations of neutrinos from the sun rules out the possibility of explaining the signal, within the thermal inert Higgs doublet we consider, if the dark matter distribution is either the SHM, or the SHM-LMC one from the simulations of~\cite{Reynoso-Cordova:2024xqz}, which does contain a high velocity tail. We note from the right middle panel that indeed, with a high enough velocity tail in the distribution, the inert Higgs doublet could explain the signal. As mentioned above, this MW-LMC-extrahigh dark matter distribution is custom made and does not reflect any currently known simulation~\cite{OHare:2026nqi,Smith-Orlik:2023kyl,Folsom:2025lly}\footnote{Considering a Maxwell-Boltzmann distribution with velocity dispersion $v_0$ and escape velocity $v_{\rm esc}$ boosted to the average Earth velocity around the galaxy $v_e$, the distribution is: $f_{\rm MW-LMC-extrahigh} = 0.986 f_{\rm MB}(v_e=251.66,v_0 = 303,v_{\rm esc}=473) + 0.014 f_{\rm MB}(v_e=45.66,v_0 = 646,v_{\rm esc}=1000)$ all in ${\rm km/s}$. These numbers are purely phenomenological but something important is that they allow the distribution to extend up to $v_{\rm min} = 1040\,{\rm km/s}$. This is relevant in the context of Eq.~\eqref{eq:vmin_kine} and the event seen at $E_{\rm R} \simeq 250\,{\rm keV}$.}. 
In other words it basically gives the minimum extra high velocity contribution one needs to accommodate both the LZ event and the solar constraints. At the same time, given our current knowledge of the high velocity tail and the difficulty of simulating it (or constrain it observationally), it cannot be ruled out. Further dedicated numerical simulations will tell whether that high velocity population is possible or plausible, or not. We note that at $1\sigma$ the fit yields $m_{H'} = 190\pm 45\,{\rm GeV}$ and $\delta m = 430\pm 30\,{\rm keV}$ as preferred ranges for the mass and mass-splitting to explain the signal given that distribution.

We note that in the past weeks several studies have explored the inert Higgs doublet as a possible explanation of the LZ signal~\cite{Nomura:2026qyq,Wang:2026ytg,Murayama:2026apt}. Refs.~\cite{Nomura:2026qyq,Wang:2026ytg} did not identify the parameter space that could be compatible with both a signal at LZ and IceCube constraints. Ref.~\cite{Murayama:2026apt} did but considered another region of parameter space with a very heavy inert Higgs $m_{H'} \simeq 15\,{\rm TeV}$, where the inert Higgs relic density arises from very large quartic interactions with the SM Higgs. In our work, we consider that these quartics are smaller and that the relic abundance of $H'$ arises only through irreducible electroweak interactions. Quite importantly, while all references highlight the technical naturalness of $\delta m \sim 500\,{\rm keV}$, see~\cite{Arina:2009um}, the model of Ref.~\cite{Escudero:2025dei} has the specificity of predicting a small mass splitting.  As explained above, on the ground of a successful asymmetry transfer between baryons and dark matter, this splitting must be rather small, $\delta m \lesssim 5\,{\rm MeV}$\footnote{This range is such that inert Higgses would be effectively captured in white dwarfs. In this context, it would be interesting to see whether they could be used as a detector for our model~\cite{McCullough:2010ai,Hooper:2010es}.}.

\section{Collider Constraints: LHC Searches and Upcoming Sensitivity from the High Luminosity Phase}\label{sec:collider}

\begin{figure}[t]
    \centering
    \includegraphics[width=0.65\linewidth]{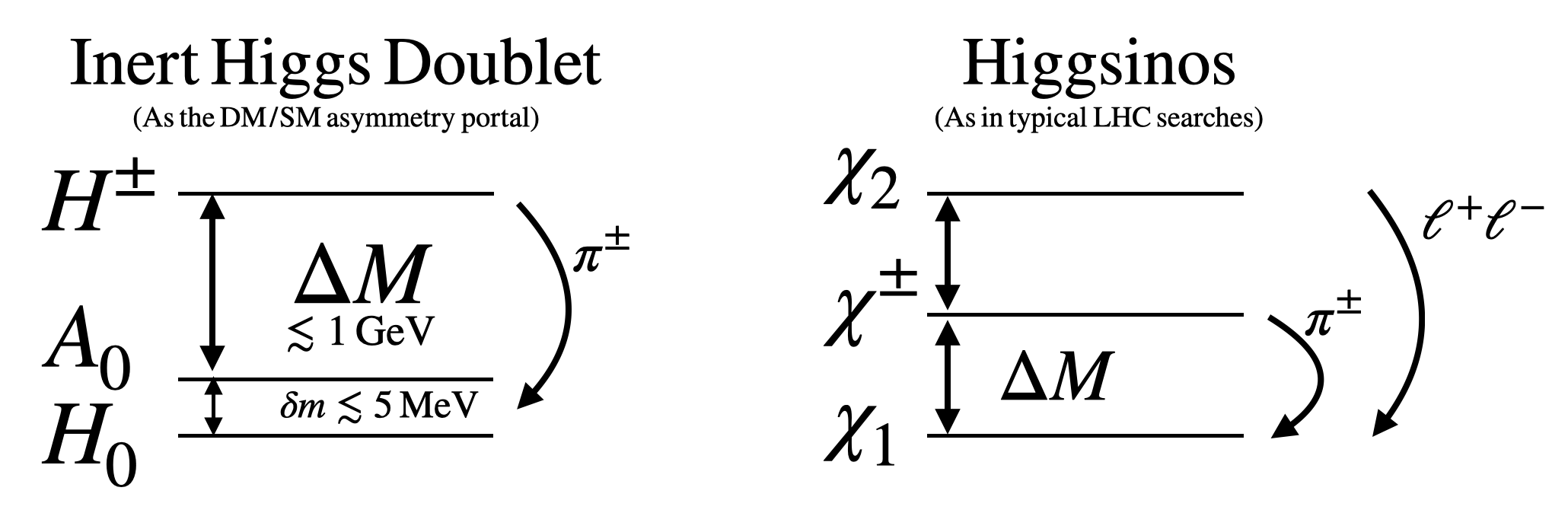}
    \vspace{-0.4cm}
    \caption{Spectrum of the compressed inert Higgs doublet, $H'$, that we consider in our study compared with the typical compressed Higgsino spectrum searched for at the LHC. The neutral splitting $\delta m\lesssim 5\,{\rm MeV}$ does not lead to relevant consequences for the collider phenomenology and we highlight the dominant decay channels for each case. In terms of collider phenomenology, the main differences between the inert doublet and the Higgsinos are: 1) that the inert Higgs has a $\sim 8$ times smaller production cross section at the LHC, 2) that as a result of the different spectrum the charged $H^\pm$ components decay twice as fast for the same given $\Delta M$, and 3) that the heavier neutral state does not decay on collider length-scales.}
    \label{fig:Hpspectrum}
\end{figure}

Collider limits on a very compressed inert Higgs doublet are in general modest, mainly due to two reasons: \emph{i}) while the doublet is produced at colliders through its electroweak charge, it mainly leads to missing energy and very soft pions arising from the decay of the heavier charged component into the neutral one -- see Fig.~\ref{fig:Hpspectrum} for a schematic on the model mass spectrum. This means that searches for this particle suffer on one hand from the need to trigger on missing energy and on the other from the difficulty to reconstruct the soft pions, a complex task given the compressed nature of the model. \emph{ii}) Being $H'$ a scalar, the relevant production rates are velocity suppressed, reducing the production cross section at colliders, when compared with its fermionic counterpart with the same quantum numbers, the Higgsino~\cite{Cirelli:2005uq}. In particular, the cross section is $\sim 30$ times smaller than for Winos and $\sim 8$ times smaller than for Higgsinos. 

The elusiveness of this compressed $H'$ is showcased by the fact that even the LEP-2 OPAL search, typically employed to constrain light and degenerate Higgsinos, was shown not capable of constraining the $H'$ for $m_\pi \lesssim \Delta M \lesssim 5\,{\rm GeV}$~\cite{OPAL:2002lje,Blinov:2015qva}\footnote{One can understand this by noting that actually not so many $W^+W^-$ pairs were produced at LEP 2, only about $\sim 10^{4}$ per experiment and, unlike $e^+e^-\to W^+W^-$, the $e^+e^-\to H'\bar{H'}$ cross section is velocity suppressed. On top of this, with such a compressed spectrum the decays from the charged components, e.g. $H^\pm\to \pi^\pm H_0$, yield little energy in the detector. The analysis to target these events requires a hard initial state radiation photon~\cite{ALEPH:2002gap} resulting in further suppression of the event yield $\sim \alpha_{\rm EM}\sim 1/129$. For the Higgsino a mass below 88 GeV could be excluded (below the kinematic threshold) but the same seems not to be true for the degenerate inert Higgs doublet. We refer to~\cite{Blinov:2015qva} for the actual reinterpretation in terms of the inert Higgs doublet which shows that no significant bound can be obtained.}. As a consequence, the strongest limit in this region of parameter space comes from the invisible decay of the $Z$ resulting in an absolute limit of $m_{H'} \gtrsim m_Z/2\simeq 46\,{\rm GeV}$. If the spectrum becomes very compressed, $\Delta M \lesssim m_\pi$, the charged scalars become long-lived and LEP analyses can bound $m_{H'}\gtrsim 95\,{\rm GeV}$ for $\tau_{H^+}> 10^{-6}\,{\rm s} $~\cite{OPAL:2003zpa}. Beyond these searches, there have been studies for the compressed inert Higgs doublet at hadron colliders considering different signatures like mono-jets, disappearing tracks and others, see for example~\cite{Belyaev:2016lok,Belyaev:2018ext,Belyaev:2020wok,Belanger:2021lwd,Belanger:2022qxt}.

In this section we explore how the LHC and its High-Luminosity (HL) phase can help probe the parameter space of a compressed inert Higgs doublet, with different analysis strategies covering different mass splitting ranges. The expected signatures are similar to those of the Higgsino, with the charged component being produced through the exchange of a photon or $Z$ in the case of $H^\pm$ pair production, or a $W$ in the case of $H^\pm$ production in association with the neutral component $H_0$ or $A_0$. Besides the smaller cross-section due to its scalar nature, the main difference in the phenomenology of the inert Higgs doublet model in regards to the Higgsino searches has to do with the mass spectrum, shown in Fig.~\ref{fig:Hpspectrum}. For the considered parameter space of the inert doublet model, the splitting between the charged and neutral components, $\Delta M$, is much larger than the splitting between the two neutral components, $\delta m$ and as such, for the small mass splittings under study, $H^\pm$ will decay to $H_0 \pi^\pm$ and $A_0 \pi^\pm$ at the same rate. This is unlike the typical Higgsino phenomenology for the LHC, where $\delta m  = 2 \Delta M$ is usually taken. Therefore the width in the inert model is twice that of the Higgsino from the extra decay channel.

Depending on the charged component decay length, see Fig.~\ref{fig:Hplifetime} for the $\tau_{H^\pm}-\Delta M$ relation, the resulting signatures we will consider are:
\begin{itemize}
    \item \textbf{Long-lived charged particles:} For $\Delta M \lesssim m_\pi$, the charged component becomes long-lived and can traverse the entire detector and hence will be targeted by heavy stable charged particle (HSCP) searches~\cite{CMS:2024nhn}. The search in Ref.~\cite{CMS:2024nhn} with a luminosity of 101 $\mathrm{fb}^{-1}$ excludes left-handed staus with $m\lesssim600$ GeV, considering only pair-production. Being doublet scalars, as the inert Higgs doublet, this analysis therefore rules out small splittings in the parameter space relevant for our model.
    \item \textbf{Disappearing charged tracks:} When the mass splitting is relatively small, $m_\pi \lesssim \Delta M \lesssim 0.3\,\mathrm{GeV}$, the $H^\pm$ will look as a long charged track in which the associated pion is too soft to be reconstructed. This results in the commonly named disappearing track signature, where a charged track stops leaving traces in the detector at some point. This track must have some minimum length to be reconstructed by the analysis~\cite{CMS:2023mny}. For reference, the innermost layers of the CMS and ATLAS pixel detectors are between $\sim 3$ and $15\,{\rm cm}$ away from the interaction point and typically these particles have boosts $\gamma \simeq 2-3$. Hence only relatively long-lived particles will hit this part of the detector before disappearing. 
    \item \textbf{Soft displaced pions:}  When the mass splitting is such that the resulting pion, despite being soft can be reconstructed, $0.3\, \mathrm{GeV} \lesssim \Delta M \lesssim 1\,\mathrm{GeV} $, one can look for soft displaced pions~\cite{ATLAS:2024umc}, taking advantage of the non-negligible decay length of the charged scalar.
    \item \textbf{Jets + missing energy:}  For mass splittings above 1 GeV, the decay products are mostly prompt and therefore the track-based searches cannot reconstruct the charged track or use the pion displacement. In this regime, one is left with mono-jet searches~\cite{CMS:2021far}.
\end{itemize}

The experimental analyses considered were primarily tailored to target Higgsinos and the collaborations provided the reach for it. Therefore, we recast each of these searches for the inert Higgs model, validating our implementations by reproducing the provided bounds on degenerate Higgsinos.

\subsection{Lifetime and branching ratio of the charged Higgs}

\begin{figure}[t]
    \centering
    \includegraphics[width=0.495\linewidth]{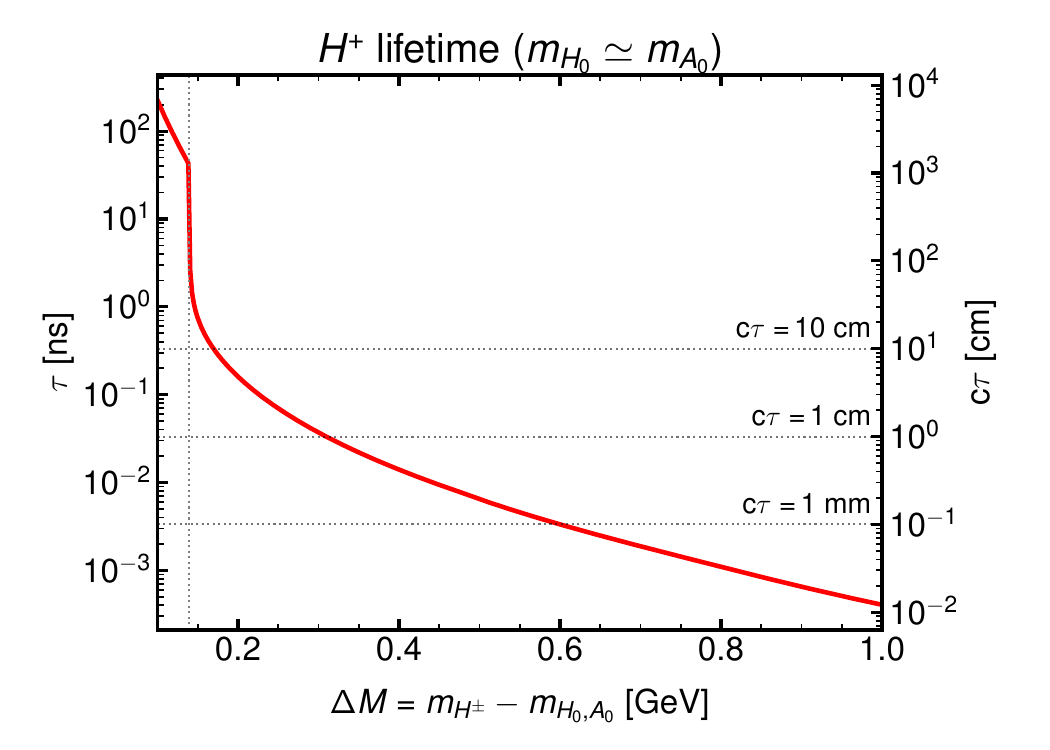}
    \includegraphics[width=0.495\linewidth]{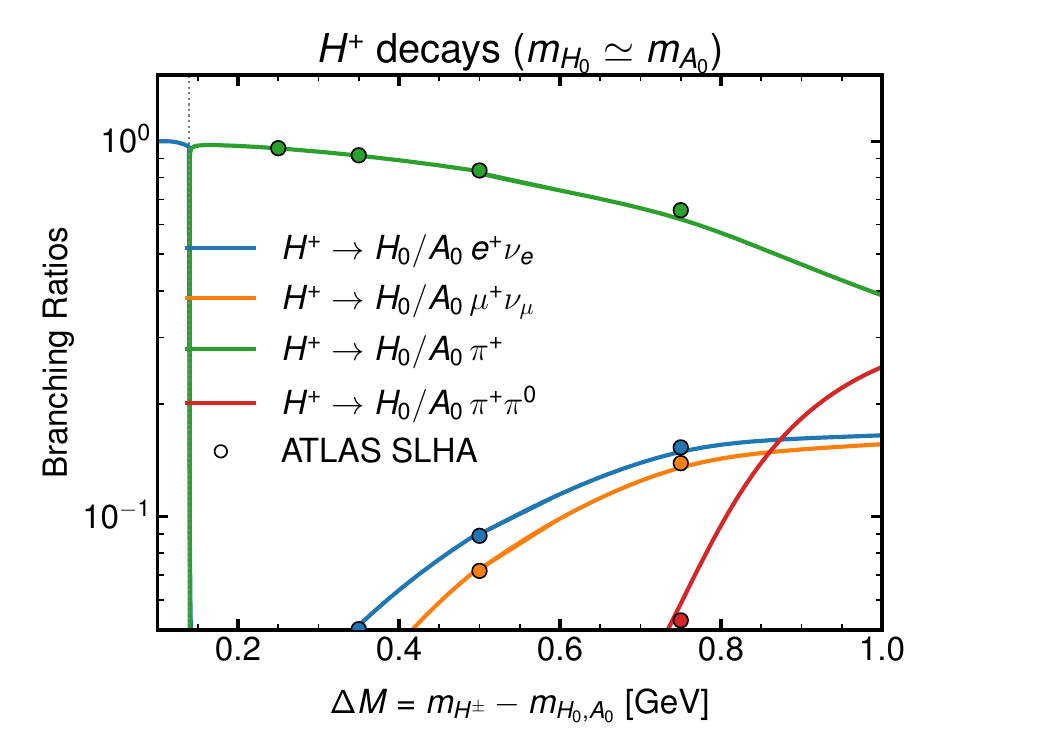}
    \vspace{-0.9cm}
    \caption{\textit{Left:} lifetime and decay length of the charged component of inert doublet as a function of the charged mass splitting, as adapted to our model from the calculation of~\cite{Ibe:2023dcu}. \textit{Right:} Branching ratios as a function of the mass splitting~\cite{Ibe:2023dcu}. The dots indicate the branching ratios taken from the ATLAS cards associated with the analysis of Ref.~\cite{ATLAS:2024umc} in the context of Higgsinos. Note that these ratios are largely independent of the overall mass scale of the inert Higgs doublet~\cite{Belyaev:2016lok,Belanger:2021lwd}. The vertical dotted line in the left plot corresponds to the charged pion mass. }
    \label{fig:Hplifetime}
\end{figure}

Fig.~\ref{fig:Hplifetime} shows the lifetime of the charged $H^\pm$ components as a function of the mass difference $\Delta M$ as computed in Ref.~\cite{Ibe:2023dcu} for Higgsinos and adapted to our inert Higgs doublet model. In the right panel one sees the relevant branching ratios. One clearly notices that as soon as the pion channel is open, it largely dominates the rates over the parameter space we consider. Its rate may be modelled via an effective vertex to reproduce the non-perturbative mixing between the $W^\pm$ and the $\pi^\pm$, $\mathcal{L}\supset g_2 f_\pi /(2\sqrt{2}) W_\mu^+\partial^\mu \pi^- + \mathrm{h.c.}$, following the studies in Refs.~\cite{Belyaev:2016lok,Belyaev:2020wok,Belanger:2021lwd,Belanger:2022qxt}. Taking into account the two decay channels ($H_0$ and $A_0$), the decay width of the charged component through the one pion channel is given, in the limit $\Delta M \ll m_{H'}$, by~\cite{Belyaev:2016lok}:
\begin{align}\label{eq:Hp_lifetime}
    \Gamma(H^\pm \to \pi^\pm H_0)+ \Gamma(H^\pm \to \pi^\pm A_0)= 2\frac{g_2^4 |V_{ud}|^2 f_\pi^2}{64\pi M_W^4} \Delta M^3  \sqrt{1-\frac{m_{\pi^\pm}^2}{\Delta M^2}}\,,
\end{align}
where $f_\pi\simeq 130\,\mathrm{MeV}$ is the pion decay constant, $V_{ud}\simeq 0.9737$, $g_2 \simeq 0.65$ is the weak gauge coupling and $M_W$ is the $W$-boson mass. This rate is independent of $m_{H'}$ up to small corrections of order $\mathcal{O}(\Delta M/m_{H'})^2 \sim 10^{-4}$. 

For mass splittings below the pion mass, the decay of the charged component is to a lepton and a neutrino, while for mass splittings above $\Delta M\gtrsim 500\,\mathrm{MeV} $ other decay channels also play a role as shown in the right hand side of Fig.~\ref{fig:Hplifetime} and Eq.~\eqref{eq:Hp_lifetime} would not serve as a fully accurate formula for the $H^\pm$ lifetime.

In all our collider analyses, we include all available decay channels by using the Higgsino signal parameter cards considered in the experimental analysis of Ref.~\cite{ATLAS:2024umc} -- available in the corresponding HEP-data repository -- after validating them and rescaling for the inert Higgs model.

\subsection{Disappearing track search for the inert Higgs doublet}

To target the disappearing track signature, we recast the CMS search of Ref.~\cite{CMS:2023mny} that used $137\,{\rm fb}^{-1}$ of data collected at $\sqrt{s} = 13\,{\rm TeV}$, see also~\cite{ATLAS:2026hnb} for a similar recent analysis by ATLAS. The CMS analysis we consider looks for charged tracks that decay within the tracker volume. Disappearing-track (DTk) candidates are divided into short or long if they contain hits recorded only in the pixel detector (short) or also in the strip detectors (long). For reference, each of these detectors extend up to $\sim 16\,{\rm cm}$ and $\sim 120\,{\rm cm}$ respectively. The track must have a transverse momentum $p_T > 25 (40)\,\mathrm{GeV}$ for the short (long) candidates, pseudo-rapidity $|\eta|<2$ and satisfy further isolation criteria, described in further detail in Ref.~\cite{CMS:2023mny}. 

The analysis includes 49 exclusive signal regions (SR). Among them, events are divided in hadronic, muon and electron channels. The latter two are not relevant for our signal and therefore we will only consider SRs 1-24. In these regions, events with one DTk candidate are split considering whether the track is long or short, through the number of jets and b-tagged jets, the ionization loss, $dE/dx$, and $p^{hard}_{T,miss}$, defined as the vector sum of the $p_T$ of all reconstructed jets with $p_T>30\,\mathrm{GeV}$ and $|\eta| < 5$. The full definition of the signal regions can be seen in Table 3 of the analysis~\cite{CMS:2023mny}.

The HEP-data repository for this analysis~\cite{hepdata.144178} provides the background yields for the SRs and therefore one needs to obtain the selection efficiencies for the signal sample only. To do this we generate the inert Higgs samples with \texttt{MadGraph5\_aMC@NLO}~\cite{Alwall:2014hca} and perform hadronization and showering with \texttt{Pythia8}~\cite{Bierlich:2022pfr}; we do not include extra partons at matrix element level reproducing the treatment of the CMS analysis for the Higgsino and Wino samples. For $m_{H'}\leq100$ GeV ($m_{H'}>100$ GeV), $10^6$ ($5\times10^5$) events were generated per mass point.

For the track reconstruction we use the short and long track reconstruction efficiencies available in the HEP-data repository~\cite{hepdata.144178}, which encode detector effects and the boosted decision tree classifier efficiency employed by CMS. These efficiencies are given as binned histograms; we tested taking the efficiency as a flat or interpolated value across the decay length corresponding to the bin and found modest effects on the final result (on the order of 10\% on the $\Delta M$ bound). We chose the flat efficiency as it was the more conservative result. The $dE/dx$ category is computed deterministically without further smearing, following Section 6 of Ref.~\cite{CMS:2023mny}. For this recast we neglect jet energy smearing and triggering efficiencies.

We validated our implementation against the Higgsino exclusions presented in the analysis and found good agreement with a maximum of $10\%$ difference in the excluded mass splitting for a given mass point.

The obtained exclusion region\footnote{We opt to shade everything below the disappearing track search as excluded. However, let us also note that for mass splittings very close to pion mass, the analysis might start losing sensitivity as tracks no longer disappear; for instance ref.~\cite{Belyaev:2020wok} found that the maximum reach was at $\Delta M \approx 155$ MeV. At these lower mass splittings, the HSCP searches will start probing the model, but there might be a $ \approx15$ MeV window ($140\,{\rm MeV}<\Delta M<155\,\mathrm{MeV}$) where the signal might evade current searches. With HL results, we expect the HSCP searches to cover this window but a careful analysis here is beyond the scope of this work.} for a compressed inert Higgs doublet are presented in the left panel of Fig.~\ref{fig:bigDelta_EWcorr_Money} whereas the projections of the reach of HL-LHC are shown in the right panel of the same figure. Collider exclusions are always shown at 95\% confidence level obtained with the CLs method~\cite{Read:2002hq}. Let us note that in~\cite{Belyaev:2020wok}, the authors recast a disappearing charged track search~\cite{ATLAS:2017oal}, and found the strongest mass bound of $m_{H'}\gtrsim 237\,{\rm GeV}$ for $\tau_{H^{\pm}}=0.53$ ns (corresponding to $\Delta M \approx 157\,{\rm MeV}$ in our setup). Our results for the same mass splitting are a bit more constraining, $m_{H'}\gtrsim 237\,{\rm GeV}$, both because the analysis we considered has more luminosity and also has a different approach, dividing the events into 49 SRs. 

From these figures, we clearly see that they cover key regions of parameter space in this model. Importantly, in the case where $\lambda_4 = 0$, then $\Delta M = \Delta M|_{\rm EW}$ and these searches are the most relevant ones on the model. For this particular coupling choice ($\lambda_4 = 0$), the current limit is $m_{H'}>170\,{\rm GeV}$ and the HL-LHC will test up to $m_{H'}\simeq 220\,{\rm GeV}$. Quite interesting, this covers well the region we identified in Section~\ref{sec:DD_inelastic} as the mass region where this particle may explain the high energy event at LZ.

\subsection{Displaced pion search for the inert Higgs doublet}

Both the ATLAS and CMS collaborations have recently released new impressive displaced pion analyses~\cite{ATLAS:2025lhc,CMS:2026ias}. However, both these recent searches are based on neural network scores for the tracks involved in each event, making the recast of this analysis unfeasible. We therefore chose to recast the simpler, cut and count based analysis from the ATLAS collaboration in Ref.~\cite{ATLAS:2024umc} which used 140 ${\rm fb}^{-1}$ at $\sqrt{s} = 13\,{\rm TeV}$. In this context, our results should therefore be taken as conservative in regards to what can be reached with current analysis techniques.

For this search, the selection criteria at the event level is given by:
\begin{itemize}[itemsep=0pt, topsep=2pt,label=--]
    \item a lepton and photon veto; 
    \item leading jet respecting $p_T > 250$ GeV and $|\eta|<2.4$;
    \item number of jets $\leq 4$; 
    \item azimuthal angle between missing transverse momentum, $p_T^{\mathrm{miss}}$, and any jet must respect $\Delta\phi(\mathrm{jet},p_T^{\mathrm{miss}})>0.4$ for all jets to suppress QCD backgrounds;
    \item missing transverse energy, $E_T^{\mathrm{miss} } = |p_T^{\mathrm{miss}}|>600$ GeV.
\end{itemize}

Within these events, at least one track must respect the following criteria:
\begin{itemize}[itemsep=0pt, topsep=2pt,label=--]
    \item $2\,\mathrm{GeV}<p_T<5\,\mathrm{GeV}$ and $|\eta|<1.5$;
    \item the transverse impact parameter of the track must respect $|d_0|<10$ mm and the longitudinal impact parameter, $z_0$, must be $|z_0 \sin \theta|< 3$ mm, where $\theta$ is the angle between the track's momenta and the beam axis; 
    \item $\Delta\phi(\mathrm{track,}E_T^{\mathrm{miss} })<0.4$; 
    \item isolation -- no other track with $p_T >1$ GeV within a cone of $\Delta R<0.4$. 
\end{itemize}

If more than one track passes the selection, the one with the maximum transverse displacement significance, $S(d_0)=d_0/\sigma(d_0)$ is considered. The resolution of the impact parameter, $\sigma(d_0)$, is $p_T$-dependent and we use the results given in ATL-IDTR-2018-008~\cite{ATL-IDTR-2018-008}. The events are then arranged into two SRs, SR-low for $8 < S(d_0) < 20$ and SR-High for $S(d_0)>20$. 

The analysis provides the background yields and therefore one only has to compute the signal efficiencies. Samples are once again simulated with \texttt{MadGraph5\_aMC@NLO}~\cite{Alwall:2014hca} and hadronization and showering is performed with \texttt{Pythia8}~\cite{Bierlich:2022pfr}. The signal samples are generated with up to 2 extra partons at matrix-element level and merged following the MLM algorithm~\cite{Mangano:2006rw,Alwall:2007fs}. Given that the analysis requires the leading jet to have $p_T>250$ GeV, we perform a generator-level cut of at least one jet with $p_T>210$ GeV. We tested that this cut did not affect the number of events passing the relevant selection. 

Our implementation was validated against the efficiencies provided in HEP-data repository~\cite{hepdata.146944} for different mass and mass splitting points for the Higgsino finding reasonable agreement in both SRs. 

The obtained bounds for the inert Higgs doublet model are shown in the left panel of Fig.~\ref{fig:bigDelta_EWcorr_Money} and the HL projection on the right panel, both denoted as displaced-$\pi$. We note that while current limits are rather weak, the HL-LHC projection from these types of searches is really powerful. In particular, it tests regions of parameter space where also direct detection experiments will be sensitive to and hence can be a critical channel in the presence of a future detection. With this, we notice that these searches may reach a sensitivity to inert Higgs doublets as heavy as $m_{H'} \sim 128\,{\rm GeV} $.

\subsection{Mono-jet recast for the inert Higgs doublet}\label{sec:mono-jet}

Jets and missing energy are the most obvious signals from a compressed inert Higgs doublet at the LHC. However, this global final state is very common in the Standard Model as well and this makes this signature on its own a somewhat complicated channel in which to search for the inert Higgs, see Fig.~\ref{fig:mono-jet} in Appendix for a clear illustration of this fact. In our model, mono-jet searches will always target production of $H_0/A_0$ and also $H^\pm$ when the charged component of the inert $H'$ are prompt, i.e. for $\Delta M \gtrsim \mathrm{GeV}$. In particular, the CMS analysis of Ref.~\cite{CMS:2021far} with 101 ${\rm fb}^{-1}$ of data at $\sqrt{s} = 13\,{\rm TeV}$ has been implemented in \texttt{MadAnalysis5}~\cite{Dumont:2014tja,Conte:2018vmg,DVN/IRF7ZL_2021}. In this context, we ran our signal samples through the \texttt{MadAnalysis5} implementation to obtain a bound on $H'$ from LHC data. To validate our simulation pipeline, we confirmed the results presented in Ref.~\cite{Agin:2023yoq} which also used the \texttt{MadAnalysis5} implementation for the Higgsino.

The $H'$ samples were generated with \texttt{MadGraph5\_aMC@NLO} and \texttt{Pythia8}. Signal samples are generated with up to 2 extra partons at matrix-element level and merged following the MLM algorithm. We considered a generator-level cut on the leading jet of $p_T> 75\,{\rm GeV}$ given that the analysis requires that at least one jet with $p_T>100\,{\rm GeV}$. We find that the mono-jet searches currently do not exclude $H'$ (at 95\% confidence level) for the simulated masses $m_{H'}>50$ GeV, independently of what the value of $\Delta M$ is. As discussed in Appendix~\ref{sec:collider}, we believe that the HL-LHC could test via this signature degenerate inert Higgses up to $m_{H'} \lesssim 55$ GeV. This is within the assumption that the $H^\pm$ decay promptly and as such we conservatively consider that it will apply to $\Delta M \gtrsim 0.6$ GeV, where the decay length is $c\tau\approx1$ mm. This region is the one precisely shown in Fig.~\ref{fig:bigDelta_EWcorr_Money}. We note that these results are in disagreement with the forecast made in~\cite{Belyaev:2018ext}, which suggests that $m_{H'}$ as large as 200 GeV may be tested via mono-jet searches by the end of HL-LHC, see lower panel of their Figure 18. 

More details on this analysis, the samples generated, and the extrapolation to HL-LHC are given in Appendix~\ref{sec:LHC_appendix}.

Another important missing energy signature for light inert Higgses are the decays of the SM Higgs into them. The current constraint on the invisible width of the Higgs is ${\rm BR}(h\to {\rm inv})< 0.107$~\cite{ATLAS:2023tkt} and one expects sensitivities of $\simeq 2.5\% $ by the end of HL-LHC. We display the current constraint on the effective inert Higgs-SM Higgs coupling in the top-left panel of Fig.~\ref{fig:elastic_results}. The limit is weak enough such that it does not appear within the $\Delta M$ figures above. For HL-LHC, in principle this would test a bit of parameter space near $m_{H'} \sim 60\,{\rm GeV}$ and for $\Delta M \simeq 1.3\,{\rm GeV}$, which would possibly extend further the reach region beyond what pure mono-jet searches could reach, but for clarity we do not display it in the figure.

\subsection{Extrapolation of recasts for HL-LHC}

In the extrapolation of the recast bounds to the HL phase of the LHC, with integrated luminosity of $\mathcal{L}=3\,\mathrm{ab}^{-1}$, one needs to quantify the share of the background uncertainty that is driven by statistics or systematic errors. To this end, and in the context of the disappearing charge track search and the displaced pions, we considered the summary plot by CMS~\cite{CMS:HiggsinoDMSummary2025}. This contains the bounds and reach for these searches at HL-LHC, in the context of Higgsinos. The results obtained in these projections consider only the statistical gain and do not assume any further technical improvements at the analysis level. To be explicit, we verified, for a fixed $\Delta M$, the improvement in the mass reach of the analysis. The HL-LHC bounds quoted imply that the systematic fraction on the background events uncertainty is $\lesssim3\%$; we therefore take the assumption that the uncertainties are statistics-dominated in our HL extrapolations for these searches. We describe in Appendix~\ref{sec:LHC_appendix} the more subtle extrapolation of the mono-jet searches to HL-LHC.

\subsection{Next-to-next-generation -- FCC-ee: Direct and Indirect Searches}\label{sec:indirect_bounds}

New particles charged under the SM gauge groups will inevitably affect electroweak precision observables in a universal way at one-loop. These corrections can be captured by the oblique parameters~\cite{Barbieri:2004qk}. In the context of minimal dark matter models, it has been shown that the $\hat{W}$-parameter, which in the heavy-mass limit one can define from $\mathcal{L}\supset-\hat{W}/(2m_W^2) (D^{\mu}W^a_{\mu\nu})^2$, is the most relevant one. The resulting contribution from the inert Higgs doublet model is~\cite{Cirelli:2005uq,DiLuzio:2018jwd,Guedes:2024vuf}:
\begin{equation}
\label{eq:wparameter}
    \hat{W} = \frac{g_2^2 }{960\pi^2} \frac{m^2_W}{m^2_{H'}}\,.
\end{equation}
Since the mass splitting is not important for this observable, this provides a bound dependent only on the mass of the doublet. 
The experimental programme of the FCC-ee will measure EWPOs with great precision; indeed Ref.~\cite{Hamaguchi:2026iry} found that a Higgsino with a mass below $500$ GeV would be excluded at 2$\sigma$ by the $\hat{W}$. For the inert Higgs, for the same expected bound on $\hat{W}$, we find the more modest reach $m_{H'}\lesssim 175$ GeV at $2\sigma$, given that the matching coefficient is a factor of 8 smaller.

This relatively small value for the mass of the BSM field questions the heavy-mass limit considered in Eq.~\eqref{eq:wparameter}. Indeed, from Refs.~\cite{Greljo:2024ytg,Maura:2024zxz,Greljo:2025ggc} the $W$-parameter is going to be bounded by FCC runs above the $Z$-pole. 
However, in Ref.~\cite{DiLuzio:2018jwd} it was shown that the heavy-mass limit underestimates the contribution to the $W$-parameter, as considering the full form factor is larger. Regardless, the mass limit is quite close to the $t\bar{t}$ run threshold, and therefore, given the large luminosity expected at FCC-ee (and despite the small production cross section of the inert Higgs doublet), one can expect direct searches to cover this mass region. For this reason, we take $m_{H'}\lesssim 175$ GeV as indicative of the FCC-ee reach. These results, and the relevance of this parameter space, therefore warrant a more detailed study of these compressed spectrum states at FCC-ee (or other lepton colliders). 

Lastly, the FCC-ee will have tremendous sensitivity to invisible decays of the Higgs reaching ${\rm BR}(h\rightarrow \mathrm{invisibles}) \lesssim 5.5\times10^{-4}  $~\cite{Selvaggi:2025kmd}. A search with such sensitivity would cut deep into the $\Delta M$ region down to $\sim 0.1\,{\rm GeV}$ at $m_{H'}\simeq 50\,{\rm GeV}$. If a clear signal is discovered at the LHC and direct detection, this would be an important observable to confirm it further at FCC-ee.
\section{Global perspective: Direct Detection and the LHC as complementary key tests of asymmetric dark matter}\label{sec:global}

\begin{figure}[t]
    \centering
    \includegraphics[width=0.6\linewidth]{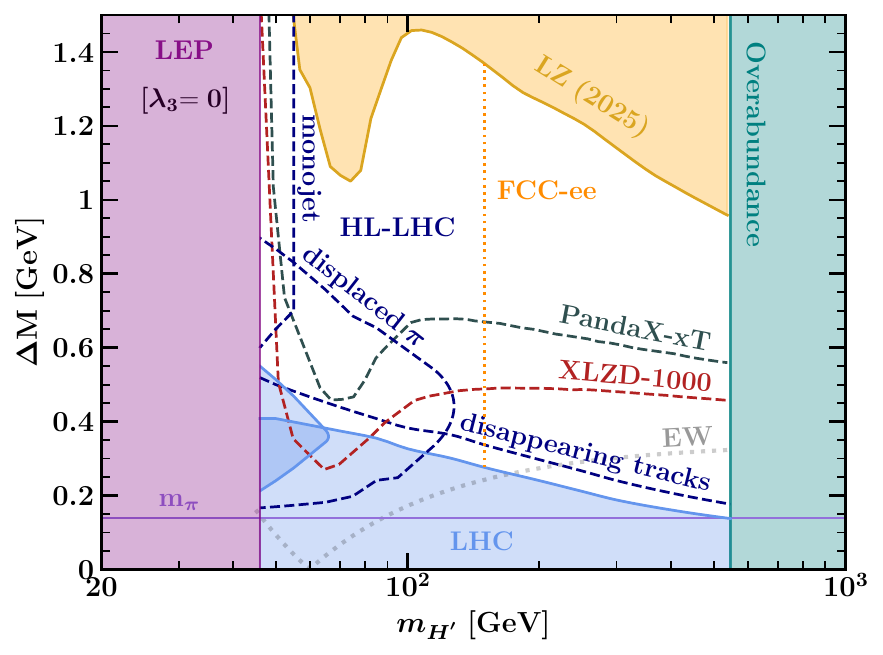}
    \vspace{-0.3cm}
    \caption{As in Fig.~\ref{fig:bigDelta_EWcorr_Money}, showing the parameter space for an inert Higgs doublet as a function of the charge-neutral mass splitting. This is for the choice $\lambda_3 = 0$ and we clearly see that while the model is relatively unconstrained at present, large regions of the parameter space will be tested by a combination of searches at direct detection experiments and at the HL-LHC. The FCC-ee line is indicative of the reach of this experiment; see text for more details.
    }
    \label{fig:Money_2_summary}
\end{figure} 

In this study we have thoroughly explored the phenomenology of a light inert Higgs doublet with $m_{H'}\leq 580\,{\rm GeV}$. This state is a well motivated particle in its own right as a possible dark matter candidate and would be 100\% of the dark matter at $m_{H'} \simeq 580\,{\rm GeV}$ or just a fraction if lighter. In this work, we have considered a light inert Higgs doublet, motivated by the asymmetric dark matter scenario recently proposed in~\cite{Escudero:2025dei}. There, the inert Higgs doublet $H'$ plays the role of the mediator which in the early Universe enables the transfer of SM asymmetries to the dark matter. In this model, the main component of the dark matter is in the form of an asymmetric population of $\phi$ particles while $H'$ is naturally only a small sub-component of the dark matter energy density. In the spirit of the asymmetric dark matter mechanism, the lighter $H'$ the more asymmetric is the dark matter. It turns out that $H'$, as a facilitator of transfers of asymmetries in the early Universe, needs to have a rather degenerate mass spectrum. From cosmological considerations, its neutral mass splitting needs to be $\delta m \lesssim 5\,{\rm MeV}$ which is technically natural. Furthermore, in the absence of cancellations, direct detection constraints require that the charged-neutral mass splitting is also small, $\Delta M \lesssim 1\,{\rm GeV}$.  Quite importantly, since $H'$ is a doublet below 580~GeV, its relic density is fixed by the Standard Model gauge interactions, which allows for very precise predictions. It is quite remarkable that this compressed inert Higgs doublet is rather unconstrained at present (it can be as light as 46 GeV) but, as we have demonstrated in our study, vast regions of its parameter space will be tested by a combination of the next generation of dark matter direct detection experiments and the HL-LHC. Some key points that we would like to address before concluding are:

\textbf{The LHC and direct detection are key complementary probes of a light compressed inert Higgs.} As stressed throughout the text, the next generation of experiments will be capable of testing most of the motivated parameter space of the model. The expected status at the end of these experiments is summarized in Fig.~\ref{fig:Money_2_summary}. The complementarity between the different probes is particularly impressive. On the one hand, collider probes, through track-based searches, will test small mass splittings, covering the parameter space from \emph{below}. On the other hand, direct detection will probe higher mass splittings closing the allowed region from \emph{above}, resulting in a very small area that is left unconstrained. Note also that several regions will be simultaneously tested by completely different probes, allowing for a discovery through one observable to be confirmed by a distinct experiment.

\textbf{The HL-LHC will be critical in testing the inert Higgs doublet:} It is often thought that the HL-LHC will not improve enough the sensitivities to new physics for a discovery to be very likely. Obviously this discussion highly depends on the type of new physics considered. As our phenomenological study shows, the scenario we consider provides a clear and motivated example of new physics for which the HL-LHC will be key to probe a large fraction of the relevant (finite) parameter space, and testing regions where a discovery is possible. In particular for the displaced pion search, it is only with the HL-LHC statistics that it plays an important role in testing this model. Our message in this context is therefore that a compressed inert Higgs doublet that may have escaped detection at LHC, can be discovered by the HL-LHC phase.

 Our collider phenomenological study is as faithful as possible, but we encourage the ATLAS and CMS collaborations to perform dedicated searches for a light and compressed inert Higgs doublet. Improvements beyond the HL-LHC scaling -- coming from detector upgrades, trigger improvements and new techniques for data analysis -- will have clear impact in testing, or discovering, this type of models, which finds theoretical motivation by being the portal between the visible and dark matter asymmetries and whose mass is expected to lie precisely in the kinematical region which is accessible to the LHC.

\textbf{Is a sub-dominant inert Higgs doublet a good candidate to explain the high energy nuclear recoil signal seen at LZ?} Theoretically, we believe that absolutely yes. Phenomenologically, as we highlight, and as was seen in other studies exploring electroweakly interacting dark matter, if the efficiency of the high energy LZ sidebands is $\sim 100\%$ then the dark matter mass should not be too big. This motivates a subdominant dark matter component with mass $m_{\rm DM}\sim 200\,{\rm GeV}$ and in the mechanism of Ref.~\cite{Escudero:2025dei} this happens naturally. Furthermore, having a small inelastic neutral mass splitting for the inert Higgs is technically natural and furthermore for $H'$ to be an effective asymmetry transfer portal it is required to be very small $\delta m\lesssim 5\,{\rm MeV}$, see~\cite{Escudero:2025dei}. As such, a splitting $\delta m \sim 430\,{\rm keV}$ that could fit the LZ signal is therefore perfectly consistent with that prediction. Of course, as was sharply highlighted in~\cite{Pospelov:2026ewn}, these types of electroweakly interacting models suffer from strong bounds from the lack of high energy neutrinos observed from the sun (even for a dark matter sub-component). In this context, and as summarized in Fig.~\ref{fig:smallDelta}, a light inert Higgs is only able to explain the LZ signal while evading these neutrino bounds provided that there exist a high enough dark matter velocity population on the tail of the distribution surrounding us on Earth. The high velocities featured in this distribution are not seen in any simulations to the best of our knowledge, and in that context this may be considered as an unlikely possibility (there is indeed a maximum theoretical speed due to the geometry and typical velocities of the dark matter population from the MW-LMC encounter). In any case, we believe that the fact that this signal can be accounted for by a high velocity tail within this well motivated particle physics model, should encourage more detailed investigations of this high velocity region at the tail of the distribution.

Time will tell whether the LZ event is truly a dark matter signal, if elastic dark matter interactions are seen in current and upcoming experiments, or whether the LHC discovers a new fundamental particle. In this study we have investigated all these aspects in the context of a light and compressed inert Higgs doublet, and we hope that our study is useful with regards to the interpretation of signals or null searches for this state in the years to come. 

\begingroup
\makeatletter
\let\addcontentsline\@gobblethree
\section*{ACKNOWLEDGMENTS}
\makeatother
\endgroup

We would like to thank Nassim Bozorgnia for sharing with us the results of the dark matter distribution on Earth from her simulated MW-LMC analogue, Nahuel Ferreiro for sending to us the expected background rate at RES-NOVA, and Chandan Hati for previous collaboration on the model considered here. M.E.A. and G.G. are very grateful to Bryan Cardwell for enlightening discussions on the LHC capabilities to search for relatively long-lived electroweakly produced particles. We have made use of Claude and ChatGPT to produce codes and double check figures supporting this work. D.C. acknowledges support from the Spanish Agencia Estatal de Investigaci\'on through the grants PID2024-155874NB-C22 and CEX2020-001007-S, funded by MCIN/AEI/10.13039/501100011033. G.G. acknowledges funding by the Eric \& Wendy Schmidt Fund for Strategic Innovation through the CERN Next Generation Triggers project under grant agreement number SIF-2023-004.
P.G.\ is supported by the SO project CEX2023-001292-S funded by MCIU/AEI/10.13039/501100011033.  The work of T.H. is supported by the Belgian IISN convention 4.4503.15 and by the Brussels Laboratory of the
Universe - BLU-ULB. A.M. has been funded through a contract linked to fellowship LCF/BQ/PI23/11970027 from “la Caixa” Foundation (ID 100010434) and the European Union’s Horizon 2020 research and innovation programme under the Marie Skłodowska-Curie grant agreement No. 847648, is also supported
by the project PID2025-169343NB-C22 and acknowledges financial support from Grant CEX2019-000918-M funded by MCIN/AEI/10.13039/501100011033. P.G., A.M. and E.P. have received support from the European Union’s Horizon Europe programme under Marie Skłodowska-Curie Actions – Staff Exchanges (SE) grant agreement No 101086085-ASYMMETRY. 
D.C., P.G., A.M. and E.P. would like to thank CERN for its hospitality during the completion of this work. 
M.E.A. is grateful to Neal Weiner and the Center for Cosmology and Particle Physics at NYU for their kind hospitality during the last stages of the completion of this work. E.P. thanks Kavli IPMU, University of Tokyo, for their hospitality during the final stages of this work.

\vspace{1cm}

\appendix
\section{Direct Detection: Parameter reconstruction and heavy targets for inelastic DM}\label{app:DD}

\begin{figure}[t]
    \centering
    \includegraphics[width=0.47\linewidth]{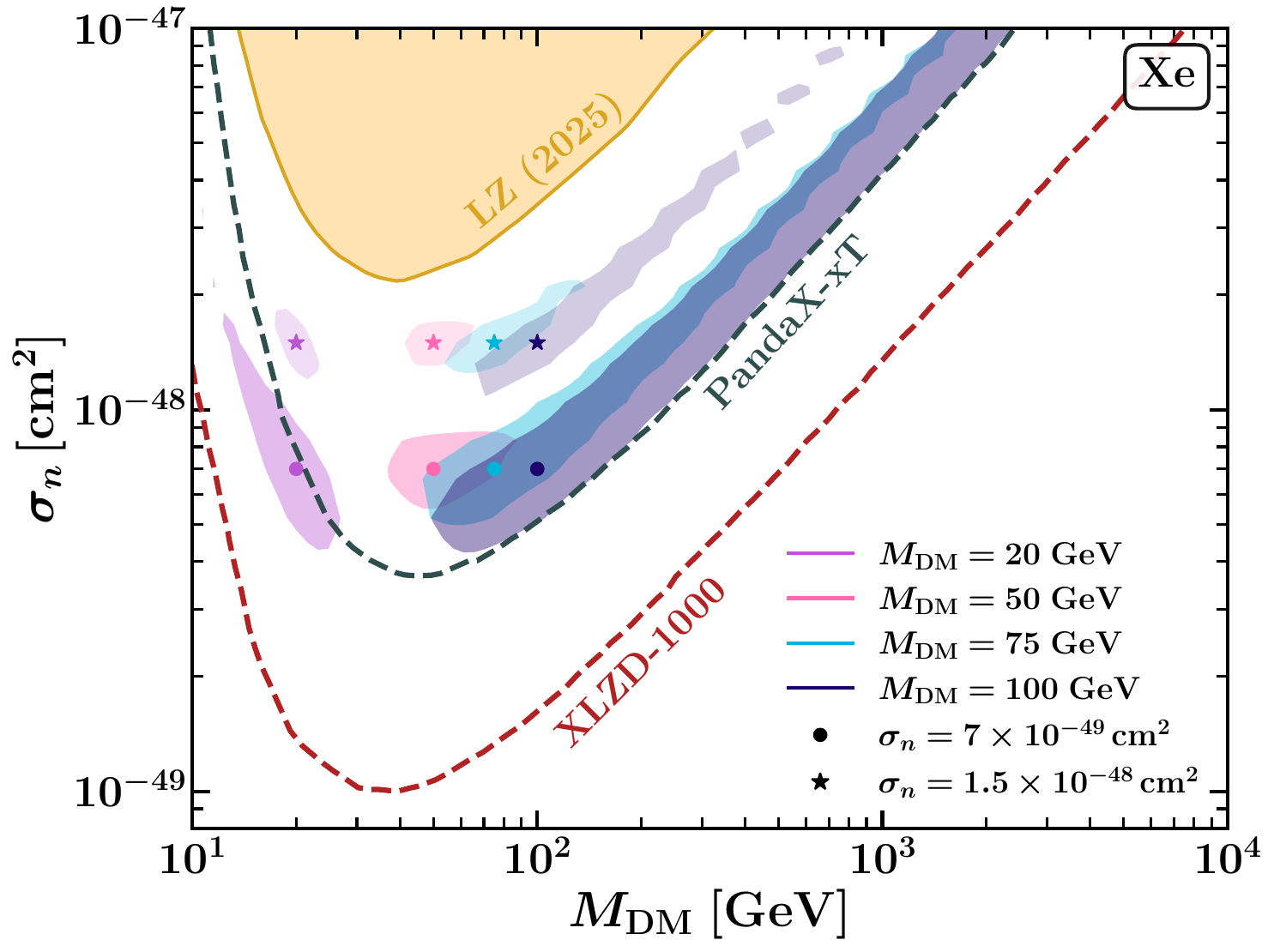}
    \includegraphics[width=0.47\linewidth]{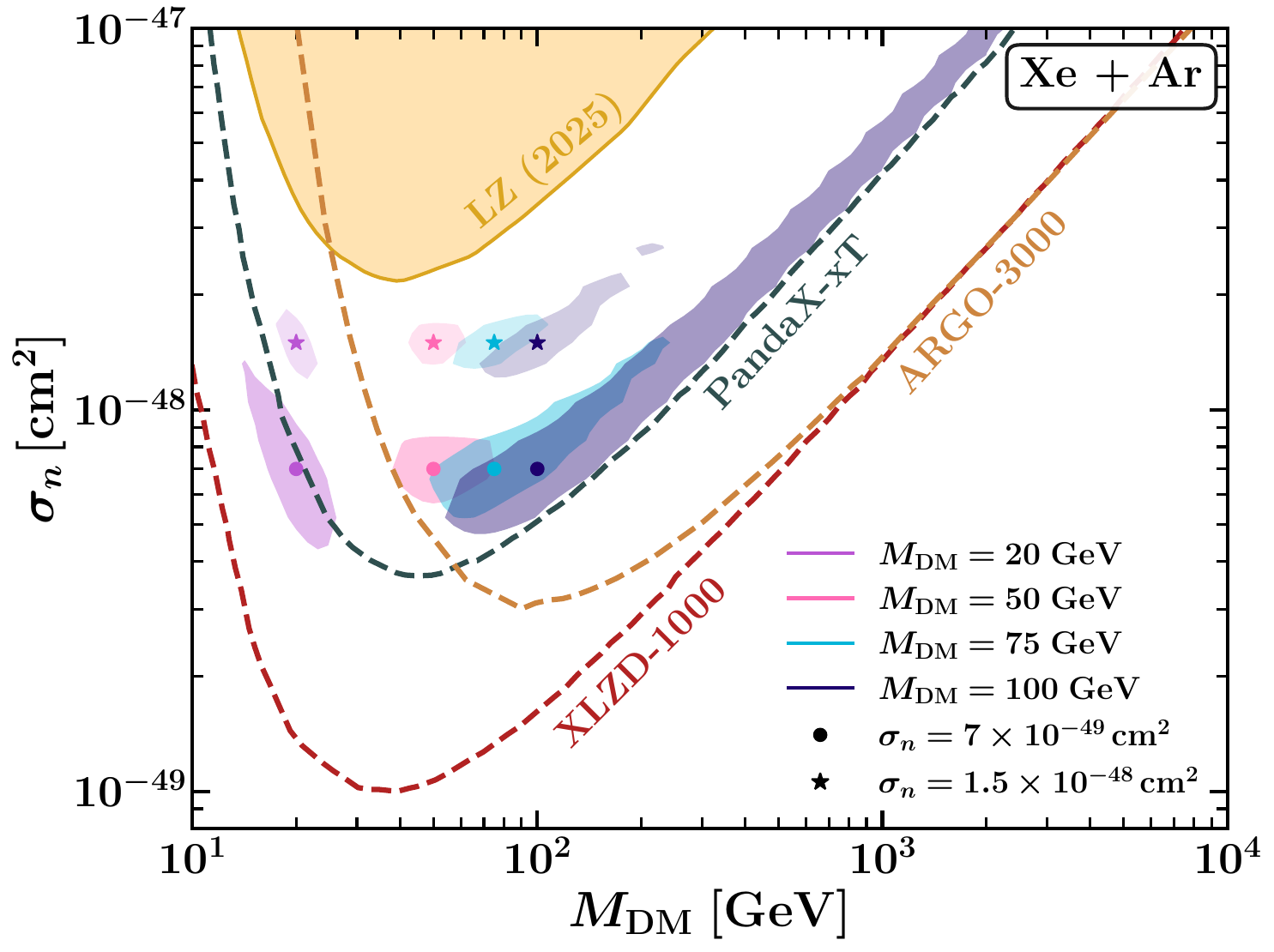}
    \caption{Reconstruction of the spin-independent elastic scattering cross section between WIMPs and nucleons ($f_n = f_p$) as a function of the WIMP mass given an injected signal (marked with stars or dots). We consider here that this WIMP makes up 100$\%$ of the dark matter. The filled contours correspond to $2\sigma$ i.e. $\Delta \chi^2 = 6.18$ while the dashed sensitivities are 90\% CL, and hence $\Delta \chi^2 = 4.61$. \textit{Left:} considering the reconstruction from XLZD-1000. \textit{Right}: adding XLZD-1000 and ARGO-3000. Also shown are the published sensitivities of PandaX-xT~\cite{PANDA-X:2024dlo} with 200 tonne-year exposure, ARGO-3000~\cite{GalbiatiARGO}, and XLZD-1000~\cite{XLZD:2024nsu}. We see that the addition of ARGO is relevant is sharpening the reconstruction of the interaction rate/dark matter mass.
    }
    \label{fig:reconstruction}
\end{figure}

The right panel of Fig.~\ref{fig:reconstruction} shows the dark matter mass and cross section reconstruction capabilities of XLZD-1000 for a generic WIMP interacting elastically with nuclei in a spin-independent way. The right panel shows how these reconstruction capabilities improve when we further combine XLZD-1000 with ARGO-3000. One can readily see that for $m_{\rm DM} \lesssim  75\,{\rm GeV}$ large enough interaction rates allow to reconstruct both the mass and scattering rates simultaneously. On the other hand, for $m_{\rm DM} \gtrsim  75\,{\rm GeV}$, lifting the degeneracy between both quantities does not appear to be feasible. Importantly, we do see that ARGO-3000 can help in this job allowing various contours to further close. In order to obtain these results, we created our own versions of the ARGO-3000 and XLZD-1000 likelihoods. We did not consider an energy smearing for simplicity. We obtained $90\%$ CL sensitivities for $\sigma_{\rm SI}$ within a factor of $\sim 30\%$ across a wide range of dark matter masses, except for $m_{\rm DM}\lesssim 100\,{\rm GeV} $ for ARGO-3000 for which we found slightly stronger limits.

\begin{figure}[t]
    \centering
\includegraphics[width=\linewidth]{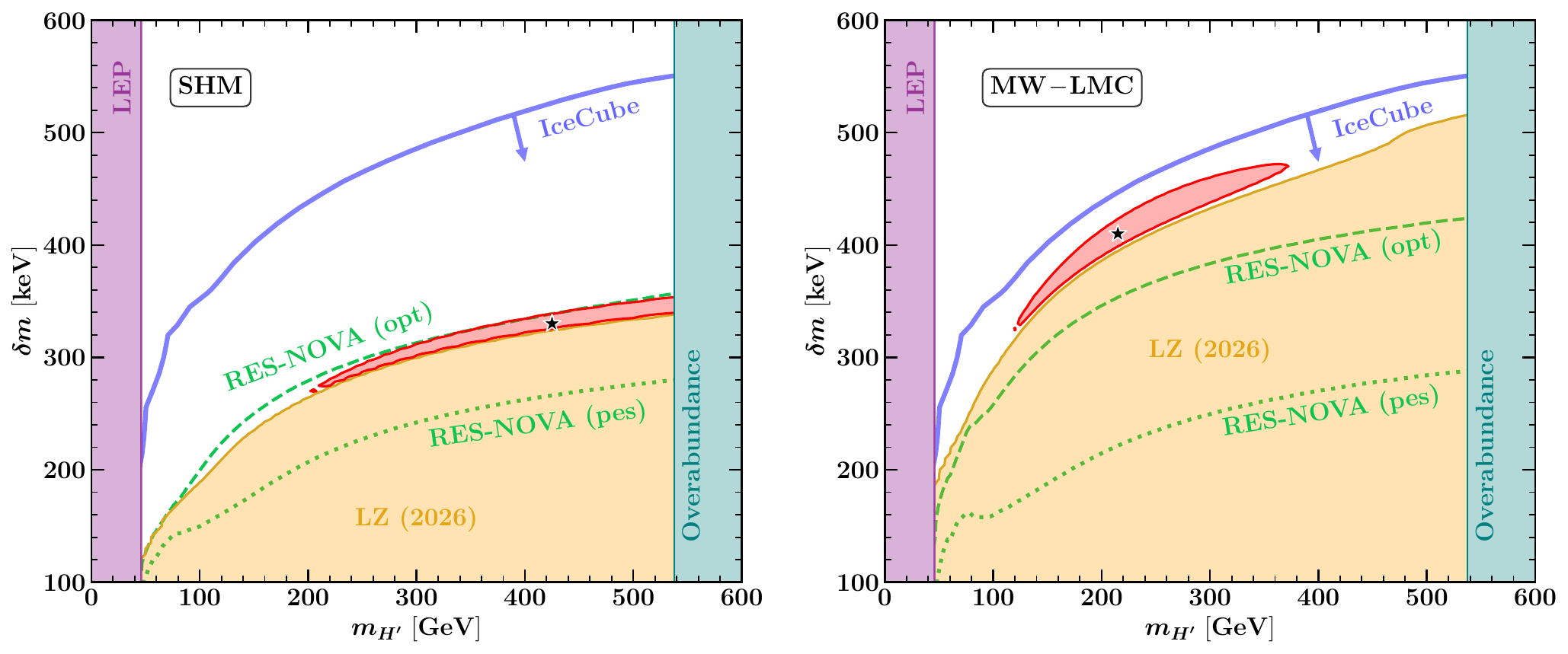}
    \caption{As in Fig.~\ref{fig:smallDelta} but including the reach from the RES-NOVA experiment with a default exposure of 170 kg-year~\cite{Res-Nova}. One sees that in the optimistic mode it may test parameter space not probed yet by LZ but only for the SHM. If there is a high velocity component in the local dark matter distribution its reach would be compromised. For $\epsilon \simeq 5\,{\rm tonne-year}$ they could reach uncharted parameter space also within the assumption of the MW-LMC dark matter distribution.
    }
    \label{fig:RESNOVA}
\end{figure}

In Fig.~\ref{fig:RESNOVA} we show the reach of the RES-NOVA experiment to the inelastic parameter $\delta m$ and compare it with the current constraints from LZ. We considered two cases: optimistic (where electronic recoils can be removed with 99.9\% efficiency) and pessimistic (where they are indistinguishable from nuclear ones). We see that only in the context of the SHM this experiment could, with its default exposure of $\epsilon = 170$ kg-year, reach a better sensitivity than what currently LZ has reached. On the other hand, if there is a high energy velocity tail, xenon experiments dominate and, even with large amounts of target mass, a RES-NOVA-like experiment would not be competitively sensitive to an inelastic signal. These conclusions can of course change if the target mass of a RES-NOVA-like experiment increases. For example, considering the MW-LMC profile, if their exposure gets to $\epsilon \simeq 5\,{\rm tonne-year}$ and adopting an optimistic configuration, they would be sensitive to yet uncharted parameter space. 

\section{Collider phenomenology: EW corrections, Higgs decays, mono-jet, and kinematics}\label{sec:LHC_appendix}
In this Appendix, we provide additional relevant details regarding our collider analysis in Section~\ref{sec:collider}. We discuss the radiative corrections to $\Delta M$, Higgs decays, the mono-jet signature, and kinematics of the inert Higgs doublet.

\textbf{Electroweak Mass splitting between the $H^\pm$ and $H_0$ masses:} Electroweak corrections naturally give a one-loop mass splitting between the charged and neutral components of the inert Higgs doublet. Importantly, as studied in detail in the main body of the paper, the $\lambda_4$ coupling gives a mass splitting already at the tree level and accordingly also at one-loop. We take the tree and loop level contributions as computed in Ref.~\cite{Cirelli:2005uq} to be:
\begin{align}
m_{H^\pm}-m_{H_0}
= - \frac{\lambda_4}{4} \frac{v_H^2}{m_{H_0}} + 
\frac{\alpha_{\rm EM}}{4\pi}\, m_{H_0} F\!\left(\frac{m_Z}{m_{H_0}}\right)\,,
\end{align}
where $F(r) =-\frac{r}{4}\left[2r^3\ln r-k r+\left(r^2-4\right)^{3/2}\ln \frac{r^2-2-r\sqrt{r^2-4}}{2}\right]$ and $\alpha_{\rm EM}\simeq 1/129$ is the fine structure constant at the $Z$ pole. Critically, $k$ in the function $F(r)$ is a UV divergent quantity~\cite{Cirelli:2005uq} but by performing the explicit one-loop calculation of the $\lambda_4$ contribution to the mass, we can compute what $k$ should be. We calculate the tadpole contribution to self-energy of the $H_0$ field, regularize it using dimensional regularization, and consider the $\overline{\rm MS}$ scheme. Mapping the result to the formula above one then finds:
\begin{align}
    k &= - \frac{\lambda_4}{4\pi \alpha_{\rm EM}} \left(\frac{m_h}{m_Z}\right)^2 \left(1+\log\frac{\bar{\mu}^2}{m_h^2}\right)\simeq -0.135 \left(\frac{\Delta M|_{\lambda_4}}{1\,{\rm GeV}}\right)\left(\frac{m_{H'}}{100\,{\rm GeV}}\right)\,.
\end{align}
We have checked that for the vast majority of the parameter space $k < 2$ and its impact on the radiatively induced $\Delta M$ is negligible as compared to the $k=0$ case. This makes sense as the one-loop contribution to $\Delta M$ from $\lambda_4$ is loop suppressed, $\sim 10^{-3}$ and hence very small. Thus, for concreteness, we simply set it to $k = 0$ in our calculations. Figs.~\ref{fig:bigDelta_EWcorr_Money},~\ref{fig:elastic_results} and~\ref{fig:Money_2_summary} shows its evolution as a function of the $H'$ mass.

\textbf{Invisible Higgs boson decay width}: The  $\lambda_3$ and $\lambda_4$ quartic couplings can also trigger the decay $h \to H^+H^-$ and $h\to H_0H_0/A_0A_0$ provided that it is kinematically allowed, $m_{H'}\lesssim m_h/2 \simeq 62\,{\rm GeV}$. The partial decay rate reads:
\begin{align}
    \Gamma(h\to H^+H^-) &= \frac{|\lambda_3|^2}{16\pi} \frac{v_H^2}{m_h} \sqrt{1-\frac{4 m_{H^\pm}^2}{m_h^2}}\,,\\
     \Gamma(h\to H_0H_0) &= \Gamma(h\to A_0A_0) = \frac{1}{2}\frac{|\lambda_3+\lambda_4|^2}{16\pi} \frac{v_H^2}{m_h} \sqrt{1-\frac{4 m_{H_0}^2}{m_h^2}}\,\quad (m_{A_0}\simeq m_{H_0})\,.
\end{align}
The branching ratio of Higgs to invisibles is constrained to be ${\rm BR}(h\to {\rm inv})< 0.107$~\cite{ATLAS:2023tkt} at 95\% CL. Assuming that a correction arises only from these new decay modes, one is able to set a limit on these new channels using the fact that $\Gamma_H^{\rm SM} \simeq 4.1\,{\rm MeV}$~\cite{ParticleDataGroup:2024cfk}. A limit is then set on:
\begin{align}
    \Gamma_{\rm inv} = \frac{{\rm BR}(h\to {\rm inv})}{1-{\rm BR}(h\to {\rm inv})} \,\Gamma_{H}^{\rm SM}\,.
\end{align}
For small $\Delta M $, one may consider that the visible decays of $H^\pm $ will be too soft to be reconstructed and hence apply the invisible limit to $\lambda_3$ in addition to the one derived for $\lambda_3+\lambda_4$ from SM Higgs decays to $H_0$ and $A_0$. We explicitly show the resulting current bound on $|\lambda_3+\lambda_4|$ in the top-left panel of Fig.~\ref{fig:elastic_results}. We see that it is weaker than current constraints from direct detection and, only for HL-LHC, it could cover a small region of parameter space at high $\Delta M$.

\begin{figure}[t]
    \centering
    \includegraphics[width=0.90\linewidth]{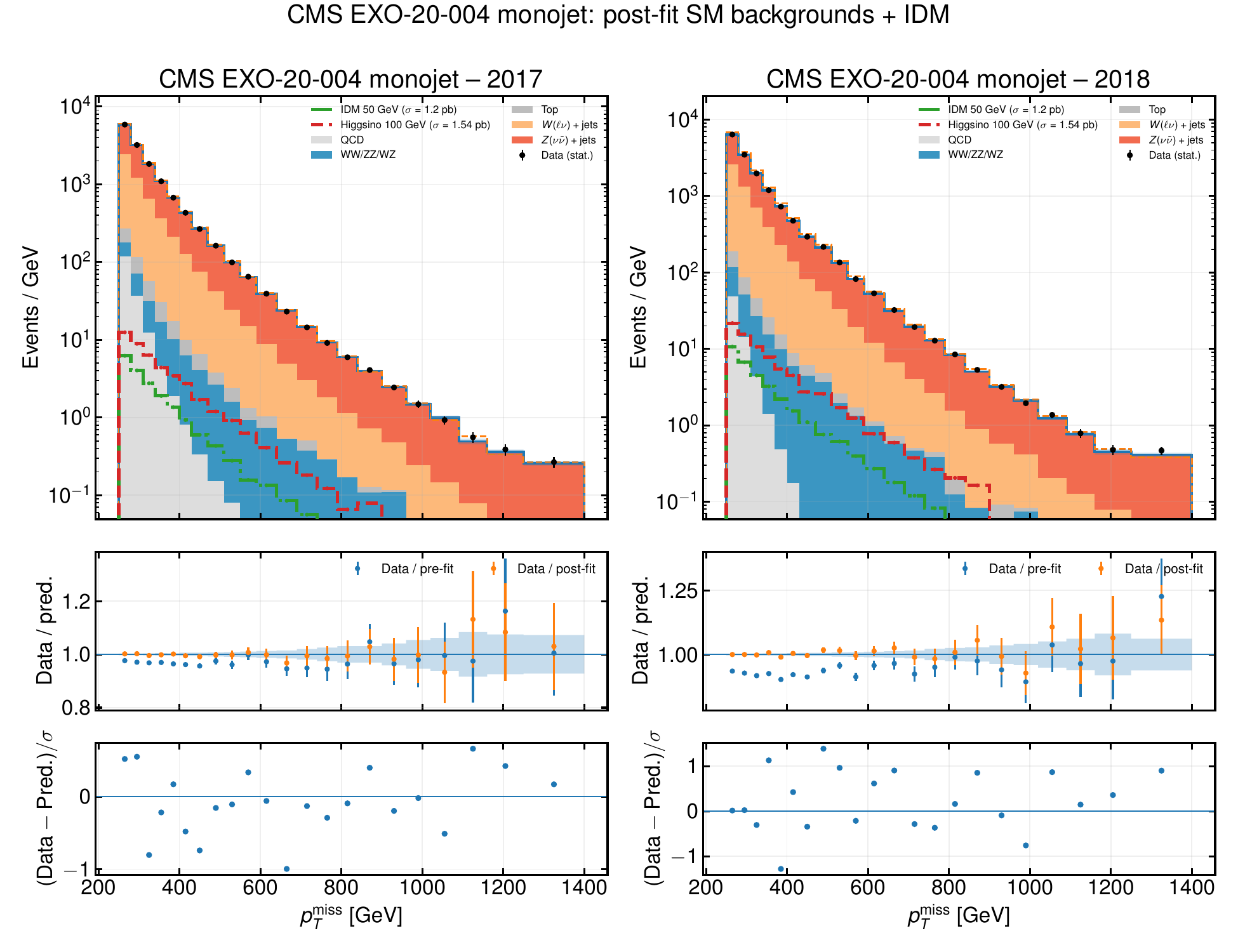}
    \vspace{-0.3cm}
    \caption{Mono-jet missing $p_T$ spectrum as in~\cite{CMS:2021far}, but with the signal from a 100 GeV Higgsino and a 50 GeV inert Higgs doublet. The 100 GeV Higgsino is currently excluded by this search~\cite{Agin:2023yoq} while the 50 GeV inert Higgs is not and may only be tested at the HL-LHC. }
    \label{fig:mono-jet}
\end{figure}

\textbf{Mono-jet at the LHC and HL-LHC:} As mentioned in Section~\ref{sec:collider}, mono-jet is a key signature of a compressed inert Higgs, but the challenge is to disentangled it from the background, that comes from various Standard Model processes leading to the same final state. This is highlighted very clearly in Fig.~\ref{fig:mono-jet} which contains the mono-jet spectrum as in~\cite{CMS:2021far} where we overlay in addition the spectrum for a degenerate inert Higgs doublet with $m_{H'} = 50\,{\rm GeV}$ and that of a 100 GeV Higgsino. One sees that the spectrum does not extend to high $p_T^{\rm miss}$ and also that it can be smaller by factors of $100$ or more as compared to backgrounds. 

The quantitative analyses of the \texttt{MadAnalysis5} recast from~\cite{CMS:2021far} is displayed in Table~\ref{tab:metjet_summary}. The table clearly highlights that 1) there is no current limit from mono-jets alone on the inert Higgs doublet, and 2) that even in the HL-LHC phase this can be rather challenging as well. As we clearly see from Fig.~\ref{fig:mono-jet}, the signal is small when compared with the large background. To estimate the reach of the HL-LHC in the mono-jet search we took the background covariance matrix provided by \texttt{MadAnalysis5}, divided the uncertainty into systematic, whose relative size was kept, and statistical. The latter was treated as uncorrelated between bins and estimated with the number of events in the control regions from where the background events were predicted from, which largely benefit from the luminosity scaling; this effect was more important in the tails of the distribution and therefore there is an improvement in the reach of the analysis. Knowing that our forecast is rather simplified, we consider that the HL-LHC may be sensitive to values of $m_{H'}$ up to $ \sim 55\,{\rm GeV}$ for $0.6\,{\rm GeV}\lesssim \Delta M \lesssim 2\,{\rm GeV}$ and we display this reach in the figures.

\begin{table*}[t]
    \centering
    \resizebox{0.65\textwidth}{!}{%
    \begin{tabular}{l||c|c|c|c|c|c}
        \toprule \hline \hline 
        & \multicolumn{6}{c}{Mono-jet Signatures of the Inert Higgs Doublet} \\
        \cmidrule(lr){2-7}\hline \hline 
        Mass [GeV] & 50 & 60 & 70 & 80 & 90 & 100 \\ \hline 
        \midrule
        Merged sample
            & $9\times 10^5$ & $9\times 10^5$  & $9\times 10^5$  & $9\times 10^5$  & $9\times 10^5$  & $4\times 10^5$  \\ \hline 
        LO prediction 13 TeV (${\rm pb}$)
            & 1.20 & 0.64 & 0.40 & 0.27 & 0.20 & 0.14 \\ \hline 
        EXP 95\%CL CMS\_EXO\_20\_004
            & 3.42 & 2.11 & 1.70 & 1.28 & 1.08 & 0.84 \\ \hline 
        OBS 95\%CL CMS\_EXO\_20\_004
            & 3.70 & 2.35 & 1.95 & 1.48 & 1.28 & 0.92 \\ \hline 
        LHC: Ratio OBS/LO
            & 3.07 & 3.65 & 4.82 & 5.40 & 7.59 & 6.34 \\ \hline 
        $K$ factor
            & 1.30 & 1.30 & 1.30 & 1.30 & 1.30 & 1.30 \\ \hline 

        \rowcolor{green!15}
        LHC: Ratio OBS/NLO
            & 2.36 & 2.81 & 3.71 & 4.16 & 5.84 & 4.88 \\ \hline 

        HL-3000 95\%CL
            & 1.55 & 0.99 & 0.80 & 0.61 & 0.49 & 0.42 \\ \hline 

        \rowcolor{blue!12}
        HL-LHC: Ratio EXP/NLO
            & \cellcolor{yellow!35}0.99
            & 1.18 & 1.53 & 1.70 & 1.95 & 2.24 \\ \hline \hline 

        \bottomrule
    \end{tabular}%
    }
    \caption{Summary of the mono-jet results. All cross sections are in ${\rm pb}$ and correspond to $p_{Tj}>75\,{\rm GeV}$ at the generator level and computed with a LO PDF. The setup assumes $\Delta M=2$~GeV between the charged component and the neutral states assuming no quartic couplings to the Higgs. We also consider results assuming a K-factor of $\simeq 1.3$ as found for Higgsinos~\cite{Fuks:2013vua,Fiaschi:2023tkq}. These results show that there is no current mono-jet limit on the Inert Higgs Doublet, and that the HL-LHC may be able to test an inert doublet up to $\sim 55\,{\rm GeV}$.}    \label{tab:metjet_summary}
\end{table*}

\begin{figure}[h]
  \centering
  \includegraphics[width=0.7\textwidth]{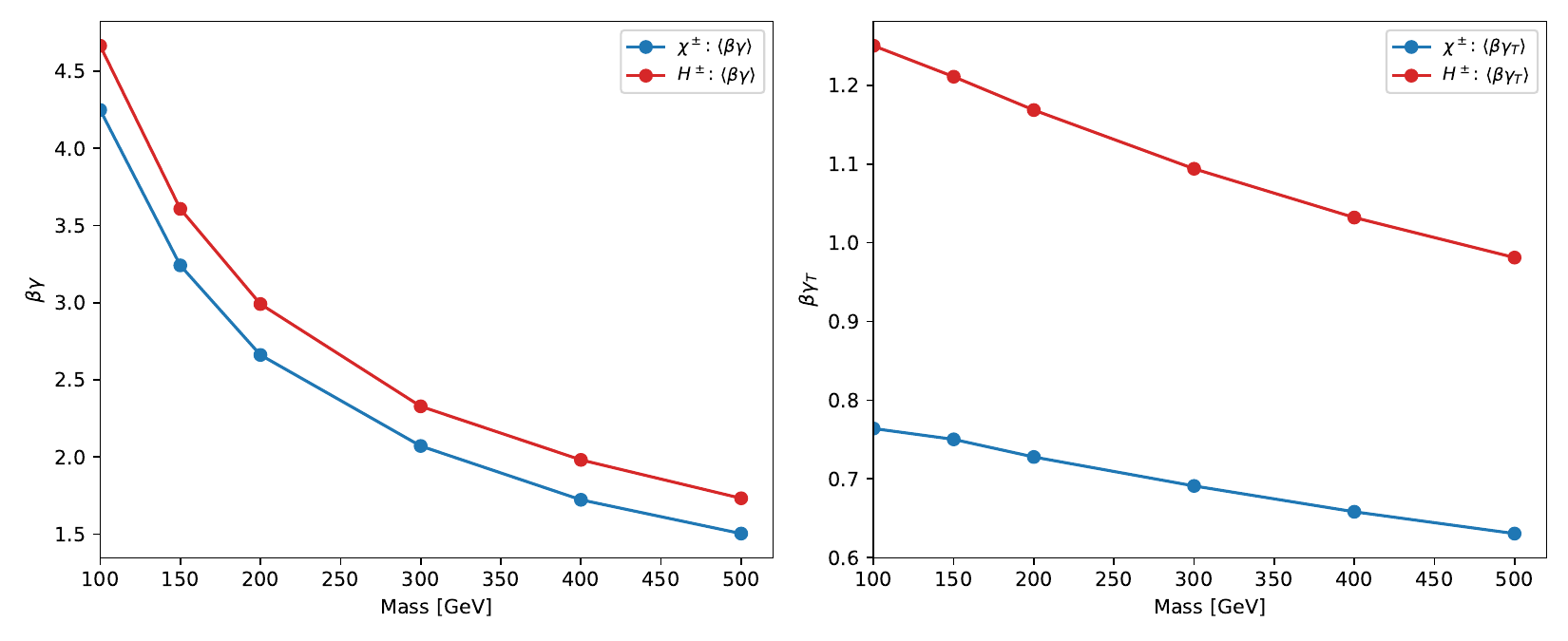}\\[0.2em]
  \begin{minipage}{0.55\textwidth}
    \centering
    \includegraphics[width=\textwidth]{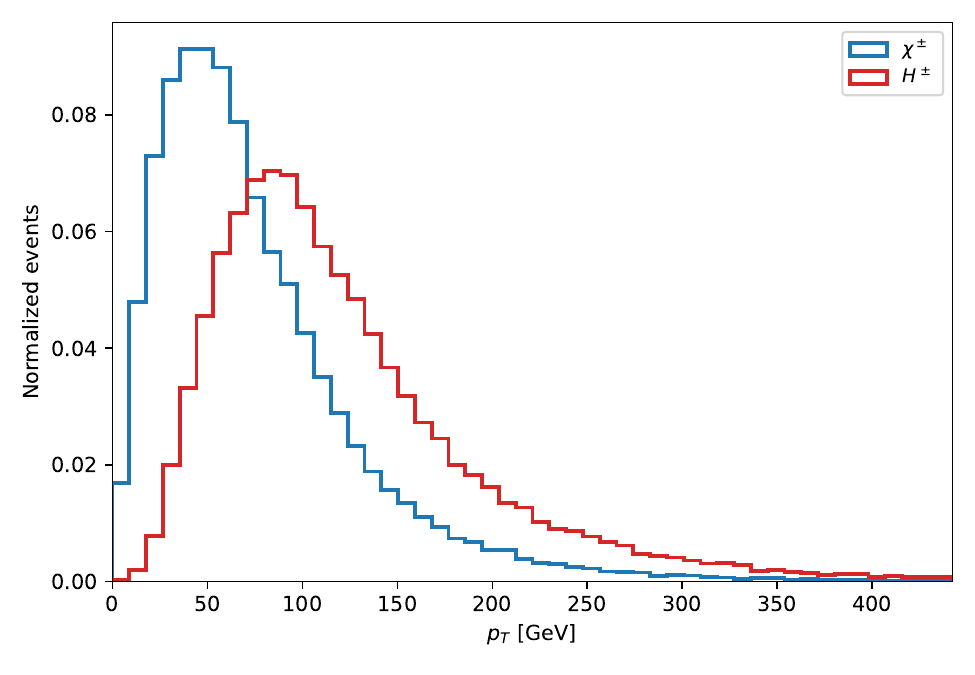}\\[0.2em]
  \end{minipage}
  \vspace{-0.4cm}
  \caption{\emph{Top Left:} Mean boost, $\beta \gamma= |p|/M$ for the chargino, $\chi^\pm$, and $H^\pm$ in the IDM. \emph{Top Right}: Mean transverse boost, $\beta_T \gamma= |p_T|/M$ of $\chi^\pm$ and $H^\pm$. \emph{Bottom:} Normalized $p_T$ distribution of $\chi^\pm$ and $H^\pm$ for a sample with mass $m=100$ GeV. The sample includes pair production and associated production with the neutral component. One notices that the inert Higgses are somewhat more boosted than their corresponding fermionic counterparts due to the velocity dependence of their production cross section.}
  \label{fig:boost_pT}
\end{figure}

\textbf{Spectrum of inert Higgses vs Higgsinos:} The recast of collider bounds from Higgsino to the inert Higgs doublet model was mainly affected by two differences: the decay length of the charged component, dictated by the mass splitting and the model mass spectrum which we have discussed in detail in Section ~\ref{sec:collider}, and the production mechanism. Regarding the latter, the inert doublet production process is velocity suppressed, significantly decreasing the cross section in comparison to the Higgsino. However, precisely due to this kinematical difference in the production process, the inert Higgs is also typically slightly more boosted than its fermionic counterpart. This is displayed in Fig.~\ref{fig:boost_pT}, which shows the boost and also the $p_T$ distribution of these two states. The differences in kinematical distributions turn out to be small, but they nevertheless do exist. It would be interesting to understand whether, in the presence of a detection from any of the channels we considered, this difference may be noticeable and lead to a model discrimination. 

\section{Bounds on the inert Higgs doublet from dark matter capture and annihilation in the sun}\label{sec:solar_appendix}

As mentioned in the main text, we have implemented the solar capture of inelastic dark matter following closely Ref.~\cite{Pospelov:2026ewn}. We have reproduced well their results. In particular, for their $1.08\,{\rm TeV}$ benchmark and in the relevant range $\delta < 550\,{\rm keV}$, the capture rate we obtain matches the capture rate of Ref.~\cite{Pospelov:2026ewn} within 20\%. Given that the capture rate spans five orders of magnitude this constitutes an excellent agreement, also because Ref.~\cite{Pospelov:2026ewn} does not provide precise information on the form factors used. In what follows, we provide some details about the implementation of these limits for the inert Higgs doublet model. 

Firstly, we consider the solar model AGSS09 from~\cite{Serenelli:2009yc}. We use the AGSS09 radial abundances directly through Ni. For the additional trace species Zn, Ge, Pb and U we adopt the central mass fractions of Table I of Ref.~\cite{DiMauro:2026dqp}. For our benchmark analysis we consider a Helm form factor of the type 
\begin{align}\label{eq:FormFactor_capture}
    F(E_R) &= 3 \frac{j_1(q\, r_n)}{q \,r_n}\,\exp \{-(q\,s)^2/2\} \,,\\
    &\quad \text{with}\quad r_n = \sqrt{c^2+\frac{7}{3}\pi^2 a^2-5s^2}\,,\,  s = 0.9\, {\rm fm}\,,\, a = 0.52 \, {\rm fm}\,,\, c = 1.23 A^{1/3} - 0.60\,{\rm fm} \nonumber\,.
\end{align}
We did test different parametrizations of the Helm form factor and found that this results in a variation of the capture rates that can be as large as $\sim 30\%$. We also explored whether the bounds changed considering a dark matter distribution with a high energy tail and found that they are rather insensitive to it.

Given the capture calculation, we consider that the inert Higgs thermalizes effectively in the sun and that therefore clumps on a typical radius around the sun of $R \simeq  0.0037 R_\odot \sqrt{1\,{\rm TeV}/m_{H'}}$~\cite{Pospelov:2026ewn}. For this to be the case, the elastic interaction cross section between the inert Higgs and nucleons should be $\sigma_{\rm SI} \gtrsim 10^{-50}\,{\rm cm}^2 $ which is fulfilled in almost all the parameter space we consider. 

\begin{figure}[t]
    \centering
    \includegraphics[width=0.47\linewidth]{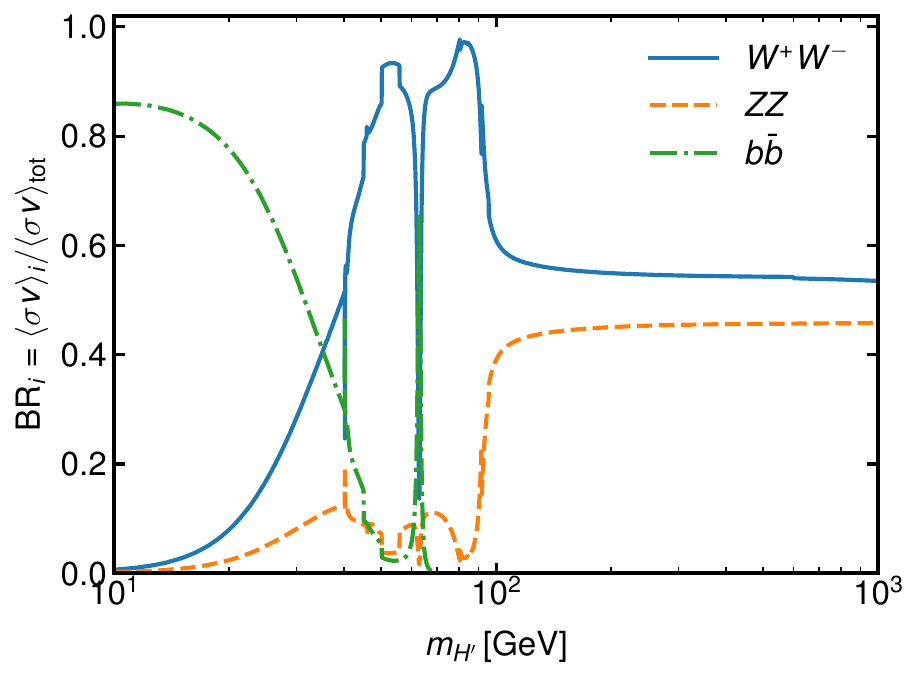}    \includegraphics[width=0.47\linewidth]{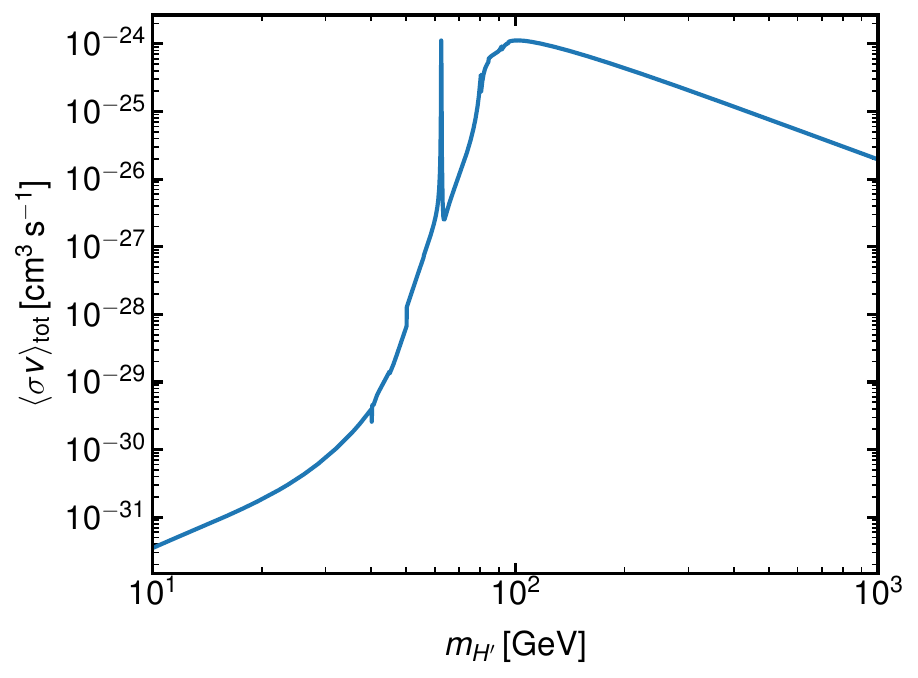}
    \vspace{-0.3cm}
    \caption{\textit{Left panel:} Branching ratios into various final states for a light inert Higgs doublet. \textit{Right panel:} Annihilation cross section evaluated at zero velocity. These results were obtained using $\lambda_3 =0$ and a quartic $\lambda_4$ coupling such that $\Delta M = 0.7\,{\rm GeV}$, but are not expected to change appreciatively for other values within the range of couplings and masses explored in our study. Computations were done using \texttt{micrOMEGAs}. Note that $W^+W^-$ and $ZZ$ final state modes are possible for $m_{H'}<m_{W,Z}$, because in this regime the code considers them as off-shell particles with multibody final states.}
    \label{fig:BR_solar}
\end{figure}

Finally, to set up a limit we consider the annihilation rate of inert Higgs doublets into $W^+W^-$ and $b\bar{b}$. For $W^+W^-$ we use the results from Fig. 2 of~\cite{Pospelov:2026ewn} and for $b\bar{b}$ we use as a limit $\Gamma_{\rm ann}^{b\bar{b}}\simeq 17\Gamma_{\rm ann}^{WW}$ as can be inferred for $m_{H'}<100\,{\rm GeV}$ from the IceCube analysis on spin-dependent dark matter capture and subsequent annihilations~\cite{IceCube:2025fcu}. In Fig.~\ref{fig:BR_solar} we show the branching ratios into the relevant final states, as well as the annihilation rate which are needed to calculate the equilibration timescale (that goes into the calculation of the annihilation rate). Comparing the calculated annihilation rate per channel with the limit from IceCube as a function of the dark matter mass we get a limit on $\delta m$ (given that the scattering cross section is fixed, see Eq.~\eqref{eq:H0inelSigma}). In the capture calculation, we take into account that $\Omega_{H'}/\Omega_{\rm DM}\leq 1$, as displayed in Fig.~\ref{fig:relicHp}. We note that, following this procedure and for a 1.08 TeV Higgsino which thermalizes in the sun, we obtain $\delta m > 560\,{\rm keV} $, in excellent agreement with the bound  $\delta m > 566\,{\rm keV}$ obtained in Ref.~\cite{Pospelov:2026ewn} (given the small differences in the solar models considered).

\bibliography{bib.bib}
\end{document}